\documentclass[twocolumn]{aastex61}
\usepackage{longtable}

\submitjournal{ApJ}

\shorttitle{Stellar Characterization of Hyades M dwarfs from the APOGEE Survey}
\shortauthors{Wanderley et al.}

\begin{document}

\title{Stellar Characterization and Radius Inflation of Hyades M dwarf stars from the APOGEE Survey}
\correspondingauthor{Fábio Wanderley}
\email{fabiowanderley@on.br}

\author[0000-0003-0697-2209]{Fábio Wanderley}
\affiliation{Observatório Nacional/MCTIC, R. Gen. José Cristino, 77, 20921-400, Rio de Janeiro, Brazil}

\author[0000-0001-6476-0576]{Katia Cunha}
\affiliation{Observatório Nacional/MCTIC, R. Gen. José Cristino, 77,  20921-400, Rio de Janeiro, Brazil}
\affiliation{Steward Observatory, University of Arizona, 933 North Cherry Avenue, Tucson, AZ 85721-0065, USA}
\affiliation{Institut d'Astrophysique de Paris, UMR7095 CNRS, Sorbonne Universit\'e, 98bis Bd. Arago, 75014 Paris, France}

\author[0000-0002-7883-5425]{Diogo Souto}
\affiliation{Departamento de F\'isica, Universidade Federal de Sergipe, Av. Marechal Rondon, S/N, 49000-000 S\~ao Crist\'ov\~ao, SE, Brazil}

\author[0000-0002-0134-2024]{Verne V. Smith}
\affiliation{NSF’s NOIRLab, 950 N. Cherry Ave. Tucson, AZ 85719 USA}
\affiliation{Institut d'Astrophysique de Paris, UMR7095 CNRS, Sorbonne Universit\'e, 98bis Bd. Arago, 75014 Paris, France}

\author{Lyra Cao}
\affiliation{Department of Astronomy, The Ohio State University, Columbus, OH 43210, USA}

\author{Marc Pinsonneault}
\affiliation{Department of Astronomy, The Ohio State University, Columbus, OH 43210, USA}

\author{C. Allende Prieto}
\affiliation{Instituto de Astrofísica de Canarias, E-38205 La Laguna, Tenerife, Spain}
\affiliation{Departamento de Astrofísica, Universidad de La Laguna, E-38206 La Laguna, Tenerife, Spain}

\author{Kevin Covey}
\affiliation{Department of Physics \& Astronomy, Western Washington University, Bellingham, WA, 98225, USA}

\author{Thomas Masseron}
\affiliation{Instituto de Astrofísica de Canarias, E-38205 La Laguna, Tenerife, Spain}
\affiliation{Departamento de Astrofísica, Universidad de La Laguna, E-38206 La Laguna, Tenerife, Spain}

\author{Ilaria Pascucci}
\affiliation{Lunar and Planetary Laboratory, The University of Arizona, Tucson, AZ 85721, USA}

\author[0000-0002-3481-9052]{Keivan G. Stassun}
\affiliation{Department of Physics and Astronomy, Vanderbilt University, 6301 Stevenson Center Ln., Nashville, TN 37235, USA}

\author{Ryan Terrien}
\affiliation{Department of Physics \& Astronomy, Carleton College, Northfield MN, 55057, USA}

\author{Galen J. Bergsten}
\affiliation{Lunar and Planetary Laboratory, The University of Arizona, Tucson, AZ 85721, USA}

\author{Dmitry Bizyaev}
\affiliation{Apache Point Observatory and New Mexico State University, Sunspot, NM 88349}
\affiliation{Sternberg Astronomical Institute, Moscow State University, Moscow, 119992, Russia}

\author{Jos\'e G. Fern\'andez-Trincado}
\affiliation{Instituto de Astronom\'ia, Universidad Cat\'olica del Norte, Av. Angamos 0610, Antofagasta, Chile}

\author[0000-0002-4912-8609]{Henrik Jönsson}
\affiliation{Materials Science and Applied Mathematics, Malmö University, SE-205 06 Malmö, Sweden}

\author{Sten Hasselquist}
\affiliation{Space Telescope Science Institute, 3700 San Martin Drive, Baltimore, MD 21218, USA}

\author[0000-0002-9771-9622]{Jon A. Holtzman}
\affiliation{New Mexico State University, Las Cruces, NM 88003, USA}

\author{Richard R. Lane}
\affiliation{Centro de Investigación en Astronomía, Universidad Bernardo O’Higgins, Avenida Viel 1497, Santiago, Chile}

\author{Suvrath Mahadevan}
\affiliation{Department of Astronomy \& Astrophysics, Pennsylvania State, 525 Davey Lab, University Park, PA 16802, USA}
\affiliation{Center for Exoplanets \& Habitable Worlds, Pennsylvania State, 525 Davey Lab, University Park, PA 16802, USA}

\author{Steven R. Majewski}
\affiliation{Department of Astronomy, University of Virginia, Charlottesville, VA 22904-4325, USA}

\author{Dante Minniti}
\affiliation{Departamento de Ciencias Físicas, Facultad de Ciencias Exactas, Universidad Andrés Bello, Fernández Concha 700, Las Condes, Santiago, Chile}
\affiliation{Vatican Observatory, Vatican City State 00120, Italy}

\author{Kaike Pan}
\affiliation{Apache Point Observatory and New Mexico State University, Sunspot, NM 88349}

\author{Javier Serna}
\affiliation{Instituto de Astronomía, Universidad Autónoma de México, Ensenada, B.C, México}

\author{Jennifer Sobeck}
\affiliation{Department of Astronomy, University of Washington, Box 351580, Seattle, WA 98195, USA}

\author{Guy S. Stringfellow}
\affiliation{Center for Astrophysics and Space Astronomy, University of Colorado, Campus Box 389, Boulder, CO 80309-0389}

\begin{abstract}

We present a spectroscopic analysis of a sample of 48 M dwarf stars ($0.2 M_{\odot}< M < 0.6 M_{\odot}$) from the Hyades open cluster using high-resolution H-band spectra from the SDSS/APOGEE survey. Our methodology adopts spectrum synthesis with LTE MARCS model atmospheres, along with the APOGEE DR17 line list, to determine effective temperatures, surface gravities, metallicities, and projected rotational velocities. 
The median metallicity obtained for the Hyades M dwarfs is [M/H]= 0.09$\pm$0.03 dex, indicating a small internal uncertainty and good agreement with optical results for Hyades red-giants. Overall, the median radii are larger than predicted by stellar models by 1.6$\pm$2.3\% and 2.4$\pm$2.3\%, relative to a MIST and DARTMOUTH isochrone, respectively. We emphasize, however, that these isochrones are different and the fractional radius inflation for the fully- and partially-convective regimes have distinct behaviors depending on the isochrone. Using a MIST isochrone there is no evidence of radius inflation for the fully convective stars, while for the partially convective M-dwarfs the radii are inflated by 2.7$\pm$2.1\%, which is in agreement with predictions from models that include magnetic fields. For the partially-convective stars, rapid-rotators present on average higher inflation levels than slow-rotators. The comparison with SPOTS isochrone models indicates that the derived M dwarf radii can be explained by accounting for stellar spots in the photosphere of the stars, with 76\% of the studied M dwarfs having up to 20\% spot coverage, and the most inflated stars with $\sim$20 -- 40\% spot coverage.
\end{abstract}


\section{Introduction}

The low-mass M dwarf stars represent a significant building block of the Milky Way, as they comprise around $70 \%$ of all stars in the Galaxy \citep{salpeter1955,reid1997,henry2016_mdwarffrac} and are thus, by number, the dominant stellar class. Another important population within the Galaxy is represented by the stellar members of open clusters, as they can be used to trace chemical evolution across the Galactic disk, and M dwarfs, due to their numbers, will also dominate the stellar census within these clusters.   
Because the stellar members of open clusters originate from the same molecular cloud, they are expected, in principle, to have nearly the same age, dynamics, and to have been born with nearly identical chemical compositions. This makes open clusters excellent stellar astrophysics laboratories in which to probe, in detail, chemical compositions across the HR diagram, including members of the M-dwarf sequence. Although the original chemical abundances of open cluster stars are expected to be nearly homogeneous, photospheric chemical variations will arise over time due to mechanisms such as diffusion that can affect main-sequence and turn-off stars \citep{souto2018a,souto2019,chaboyer1995,richard2005,dotter2017,gao2018,bertelli2018,liu2019}, or convective dredge-up in red-giant stars \citep{beckeriben1979,lagarde2012,salaris2015}, that modify their surface abundances. 
M dwarfs, on the other hand, are not expected to suffer significantly from either effect due to their convective envelopes, although changes to surface $^{3}$He and $^{4}$He abundances take place in the lower-mass fully-convective M-dwarfs, due to contact between the surface and the H-burning core.  Although He abundances in M dwarfs are not directly measurable via spectroscopy, evidence for the boundary between the partially-convective and fully-convective M dwarfs, and the role that $^{3}$He burning plays in this boundary \citep{vansaders2012}, has been revealed as a gap in the Gaia M$_{G}$ -- (G$_{BP}$ -- G$_{RP}$) diagram (e.g., \citealt{jao2018,feiden2021}).  Thus, other than He in the lower-mass M-dwarfs, these stars will carry the fingerprints of near-pristine cluster abundances and are good indicators of the true cluster chemical composition.

The modeling of M dwarfs has advanced in recent years. Several works adopting high-resolution spectrum synthesis techniques  \citep{souto2017,souto2018b,souto2020,souto2022,Lindgren2016, Lindgren2017, Veyette2017, Rajpurohit2018a, Rajpurohit2018b, Passegger2018, Passegger2019a, Passegger2019b, Passegger2020a, Passegger2020b, Lopez-Valdivia2019, Birky2020, Marfil2021, Sarmento2021, Khata2021} have demonstrated that reliable stellar parameters for M dwarfs can be obtained, although there are certainly systematic differences between the metallicity results that still need to be evaluated in detail. The study of M dwarfs in open clusters as benchmarks offers the possibility to test the metallicity scale as a function of the M-dwarf effective temperature and also quantify possible systematic metallicity differences between the M dwarfs and the hotter FGK stars.

Very few chemical abundance studies of stars in open clusters to date have analyzed M dwarfs. In recent work, \citet{souto2021} presented one of the first high-resolution spectroscopic studies of an open cluster that included a sample of M dwarfs. They used near-infrared spectra obtained by the SDSS APOGEE survey \citep{majewski2017_apogee} to derive metallicities in seven M dwarf members of the solar neighborhood (d=85 pc; \citealt{tang2018}), young (age=600 Myr; \citealt{casewell2006,casewell2014,kraushillenbrand2007,tang2018}), near-solar metallicity, open cluster Coma Berenices. The Coma Berenices M dwarfs analyzed by \citet{souto2021} covered the T$_{\rm eff}$ range between roughly 3100 and 3800 K, but also included hotter stellar members of Coma Berenices, having F, G, and K spectral types in the analysis, finding a decrease in the mean metallicities of $\sim$0.05 dex between the more massive K dwarfs relative to the M dwarfs; such a decrease in the metallicities is in line with predictions of atomic diffusion from MIST isochrone models \citep{choi2016_MIST}.
 
The Hyades cluster (Melotte 25) is another young (625 $\pm$ 50 Myr) solar neighborhood open cluster that is quite nearby (center of mass located at $\sim 43.3 \pm 0.3$ pc from the Sun \citealt{perryman1998}). 
Previous studies of the Hyades analyzed optical spectra of red-giants and dwarfs of F, G, and K types. Most of these studies found the Hyades to be slightly metal-rich relative to the Sun. \cite{heiter2014} conducted a critical compilation of literature high-resolution spectroscopic metallicity studies in open clusters and collected 129 metallicity measurements from 22 works in the literature for the Hyades cluster through 2013. After a careful examination of the studies and their uncertainties, they estimated a metallicity for a sample of 16 red-giant stars to be [Fe/H]=$+0.12 \pm 0.04$ and for a sample of 76 dwarfs as $+0.13 \pm 0.06$. 
Various other more recent works using high-resolution spectroscopy determined metallicities for samples of main-sequence stars that included Hyades members, such as \citet{Brewer2016}, \citet{Aguilera-Gomez2018}, and \citet{Takeda2020}. 
The average metallicities of these studies obtained for the Hyades stars are, respectively, +$0.22 \pm 0.02$ (for 10 solar-type stars), +$0.1 \pm 0.02$ (for 7 F-type stars) and +$0.13$ (for 5 solar-type stars), all indicating that the Hyades open cluster is metal-rich.
A cross-match of the latest version of the PASTEL catalog \citep{Soubiran2016_pastel} (Version 2020-01-30; a compilation of measurements of effective temperatures, surface gravities, and metallicities obtained from high-resolution, high signal-to-noise spectroscopy) with the list of Hyades members from \citet{goldman2013} finds hundreds of matches (these results will be discussed in Section 4). However, despite the great amount of data available in the literature for the Hyades cluster, and the general consensus about it being a metal-rich open cluster, there have been no high-resolution spectroscopic abundance studies of the Hyades M dwarfs.

Because the Hyades cluster is quite nearby, its M dwarf members can be observed with available mid-size telescopes and high-resolution spectrographs, such as those of the SDSS APOGEE survey \citep{majewski2017_apogee}. APOGEE has targeted a large number of Hyades members, including faint M dwarfs to a magnitude limit of $H \approx 12.1$. 
In this work, we use APOGEE (Apache Point Observatory Galactic Evolution Experiment) spectra to determine effective temperatures, surface gravities, and metallicities, along with stellar masses and radii for a sample of 48 M dwarf members of the Hyades open cluster. 

One aspect that can be investigated with the derived stellar radii for the APOGEE Hyades sample is whether the studied M dwarfs have inflated radii, as previous works have found evidence of radius inflation in low mass stars, including M dwarfs \citep{reiners2012,Jackson2016,Jackson2018,Jackson2019,jeffers2018,Kesseli2018,jaehnig2019}.  Radius inflation could be related to the presence of magnetic fields in M dwarfs through the inhibition of convection and/or dark surface star spots, as explored by \citet{chabrier2007}.  In the case of strong global B-fields, they found that if the fields inhibit convective efficiency, overall stellar luminosity is lowered, as well as T$_{\rm eff}$, while the radius increases. 
Dark surface star spots, as modelled by \citet{chabrier2007}, reduce the disk-integrated effective temperature and, since the stellar luminosity is unaffected, the radius of the star expands \citep{feiden2012,gough1966}.
In this paper, we will compare the derived radii with models including those that consider stellar spots.

This paper is organized as follows: Section 2 presents the APOGEE spectra analyzed, and discusses membership and sample selection. Section 3 describes the adopted methodology used in the spectral analysis. Results are presented in Section 4, and in Section 5 we discuss radius inflation in the Hyades, along with indicators of X-ray and UV activity for the studied M dwarfs. Finally, Section 6 summarizes the conclusions.

\section{APOGEE Data and Sample Selection}

In this work, we analyze spectra obtained by the SDSS IV APOGEE Survey \citep{majewski2017_apogee} (most of them obtained by data release 17, hereafter DR17 \citealt{abdurro'uf2022_dr17,blanton2017_sdss}). The Hyades open cluster was observed using the cryogenic, multi-fiber (300 fibers) APOGEE-N and APOGEE-S spectrographs on the 2.5-m telescope \citep{bowen1973,gunn2006_sdss,wilson2019} located at APO (Apache Point Observatory) and the 2.5-m du Pont Telescope at Las Campanas Observatory, respectively. The near-infrared APOGEE spectrographs operate in the H band, covering the spectral range between $\lambda$1.51$\mu$m to $\lambda$1.69$\mu$m, at an average spectral resolution of $R\sim22,500$. 

The APOGEE survey systematically targeted a large number of stellar members of open clusters, through the OCCAM (Open Cluster Chemical Abundances and Mapping) survey \citep{donor2020_occam}, with the Hyades open cluster included in an effort referred to by the APOGEE targeting team as BTX, or \textit{bright time extension}. The APOGEE BTX \citep{zazow2017,beaton2021} began in 2017 with several goals, one of which was to extend APOGEE observations, as much as possible, towards low-mass main sequence stars and to observe stars from the Kepler-2 (K2) mission  \citep{howell2014_k2} fields: the Hyades cluster was one of the fields observed by K2, making it one of the targets for BTX. 

The selection of Hyades cluster members by the APOGEE targeting team was based on the previous works of \citet{roser2011} and \citet{goldman2013}, who used kinematic and photometric stellar parameters to identify 773 stars as possible Hyades members, of which 238 stars were observed by the APOGEE survey.
We selected the APOGEE Hyades sample of stars which are included in GAIA EDR3 \citep{GaiaCollaboration2021_gaiaedr3} and have measured distances in \citet{bailerjones2021}. 

We then used the results from \citet{douglas2019} to remove those stars which were candidate members of binary or multiple systems; all stars in their list flagged as being a binary (based upon at least one of their binarity criteria) were removed from our sample. \citet{douglas2019} used K2 light curves to derive rotational periods for low-mass Hyades stars and also compiled other measurements from the literature \citep{radick1987_refdouglas1,radick1995_refdouglas2,prosser1995_refdouglas3,delorme2011_refdouglas4,hartman2011_refdouglas5,pojmanski2002_refdouglas6,douglas2016_refdouglas7}. Many stars in our sample also had their K2 light curves and rotation periods studied in \citet{stauffer2018}. Several stars had more than one rotational period measurement, both in \citet{douglas2019} and \citet{stauffer2018}, with some of them showing deviations in these measurements. A large deviation in rotational period measurements can be an indicator of binarity, although it can also be a sign of differential rotation or spots. As proposed by \citet{douglas2019}, we adopted $20 \%$ as a threshold in the variation of the rotational periods as being due to binarity, assuming that lower variations can be explained by other mechanisms. We also checked our sample for binarity using the Washington Double Star Catalogue \citep{mason2001}, as well as, \citet{ansdell2015}. In addition, those stars having very large Gaia RUWE numbers (RUWE $>$ 1.6) have not been included in our sample, as RUWE can give an indication of the presence of an unresolved companion \citep{belokurov2020}. 

\begin{figure}[h!]
\begin{center}
  \includegraphics[angle=0,width=1\linewidth,clip]{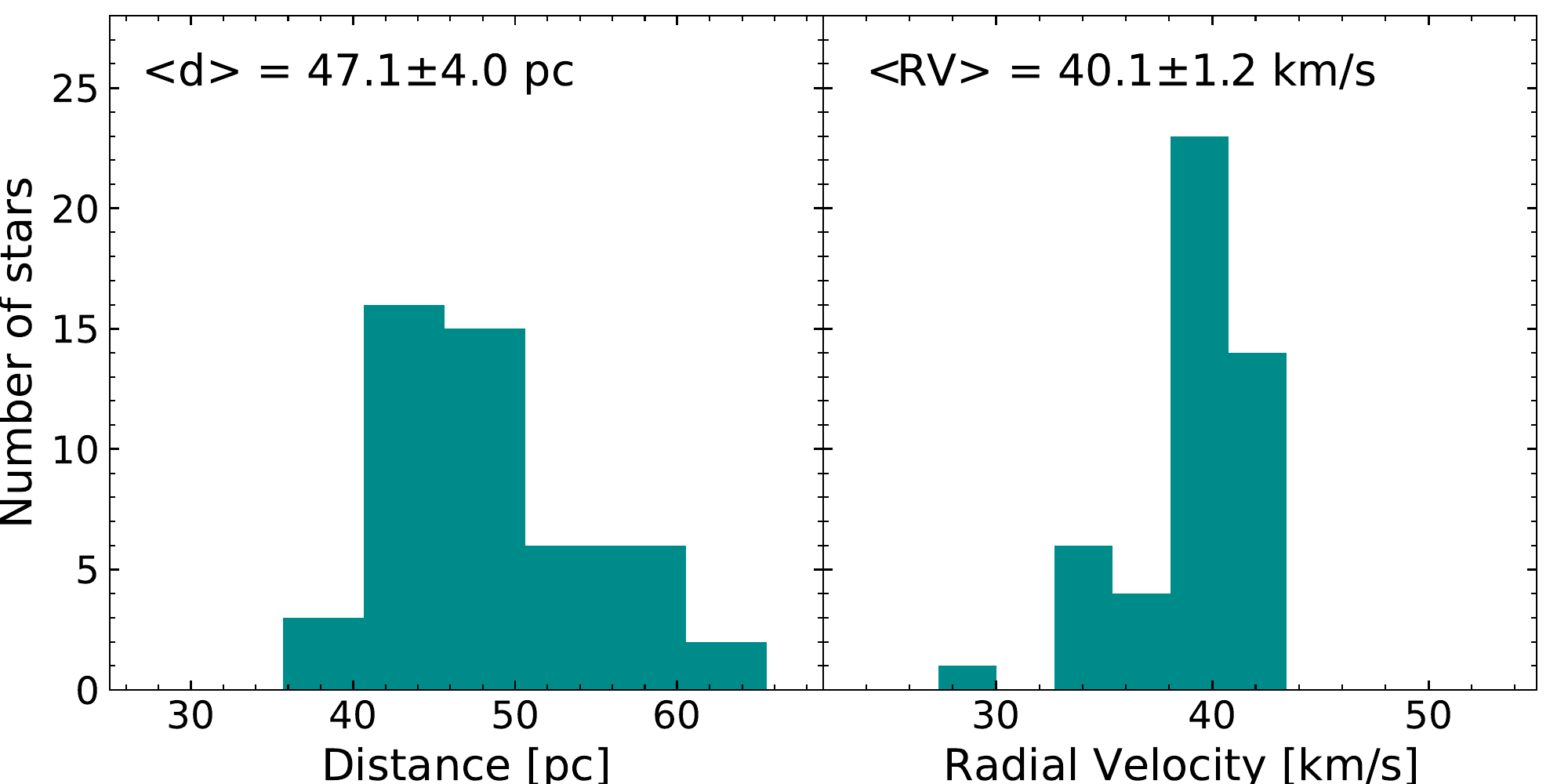}
\caption{The left and right panels present, respectively, the distribution of distances (from \citealt{bailerjones2021}) and the radial velocities measured by APOGEE for the sample of studied M dwarf stars in the Hyades open cluster.}
\end{center}
\label{membership}
\end{figure}

We also checked the sample for stability in the radial velocity values measured by APOGEE and removed those stars that showed a scatter in radial velocity greater than 1 km s$^{-1}$. Each APOGEE observation or namely \textit{visit} generates one radial velocity measurement \citep{elbadry2018}. We note, however, that many stars in our sample had only one APOGEE visit. 
Finally, we also removed from our sample those stars with outlier radial velocities (see Figure \ref{membership} for the final sample RV distribution) and proper motions.

In this study, we focus on the analysis of M dwarfs with spectral types between $\sim$M1 - M5 and these correspond to a magnitude range roughly between $11.9 < G < 15.8$ and color range between $2.1 < G_{\rm BP}-G_{\rm RP} < 3.3$ 
(\citealt{pecault_mamajek_2013};  \citealt{bentley2018}; \citealt{cifuentes2020}). The sample of Hyades M dwarfs selected to be analyzed here is composed of 48 single M dwarfs having APOGEE spectra with S/N $>$ 70.
Table \ref{stellartable} presents the photometric data for the sample stars  (2MASS; \citealt{Skrutskie2006_2mass} and Gaia passbands; \citealt{GaiaCollaboration2021_gaiaedr3,GaiaCollaboration2016_gaiamission}), distances \citep{bailerjones2021} (along with its uncertainties), the APOGEE radial velocities, and the S/N for the analyzed APOGEE spectra.

In Figure \ref{membership} we present, from left to right respectively, the distance (from \citealt{bailerjones2021}) and radial velocity distributions for the studied M dwarf stars. The mean distance obtained for the sample is $47.1 \pm 4.0$ pc, and the scatter within this sample can be explained by the cluster depth along our line of sight and is within the typically cited tidal radius of the Hyades cluster of $\pm 10$ pc \citep{reino2018}. The average RV of our sample is $40.1 \pm 1.2$ km s$^{-1}$ and this compares well with other radial velocity determinations from the literature for the Hyades, e.g., $39.36 \pm 0.26$ km s$^{-1}$ from \citealt{leao2019_rvref}.

\section{Methodology}

\subsection{Stellar Parameters and Metallicities}

We performed spectral synthesis modeling for APOGEE spectra, in order to derive effective temperatures, surface gravities, metallicities, oxygen abundances, and projected rotational velocities for 48 M dwarf stars from the Hyades open cluster. This spectral analysis employed 1-D plane-parallel LTE MARCS model atmospheres \citep{Gustafsson2008_marcs}, the APOGEE DR17 line list \citep{shetrone2015,smith2021} and the radiative transfer code Turbospectrum2020 \citep{plez2012_turbospectrum} 
to compute synthetic spectra. We used the BACCHUS \citep{masseron2016_bacchus} wrapper to derive chemical abundances, but took into account the associated line spread function (LSF) for each of the spectra analyzed, as the APOGEE spectra show fiber-to-fiber LSF variations, as well as wavelength LSF variations across the chips \citep{nidever2015,wilson2019}.

The M dwarf model spectra have low sensitivity to the microturbulent velocity parameter, with a value of 1 km s$^{-1}$ providing good fits to the observations \citep{souto2017}, and this value was adopted in all calculations.

The effective temperatures and oxygen abundances in this study were derived using the methodology presented in \citet{souto2020}, which is based upon measurements of H$_{2}$O and OH lines in APOGEE spectra of M dwarfs. The principle is simple: while the strength of the H$_{2}$O and OH lines are both sensitive to the oxygen abundance, the measured water lines are highly sensitive to $T_{\rm eff}$, while the OH lines are rather insensitive to changes in the effective temperature. 
This difference in behavior is due, in part, to the measured OH lines arising generally from lower-excitation energy levels ($\chi\sim$0.2--0.9 eV), while the measured H$_{2}$O lines are dominated by higher-excitation levels ($\chi\sim$1--2.5 eV).
In addition, the partial pressures of H$_{2}$O change much more than OH in the stellar atmosphere line-forming regions as a function of T$_{\rm eff}$ across the M-dwarf sequence. The resulting differences in the behavior of the water and OH lines as functions of the effective temperature can be used to constrain T$_{\rm eff}$ by defining a common $T_{\rm eff}$-A(O) pair that brings the oxygen abundances of OH and H$_{2}$O into an agreement. 

\begin{figure}[h!]
\begin{center}
\includegraphics[angle=0,width=1\linewidth,clip]{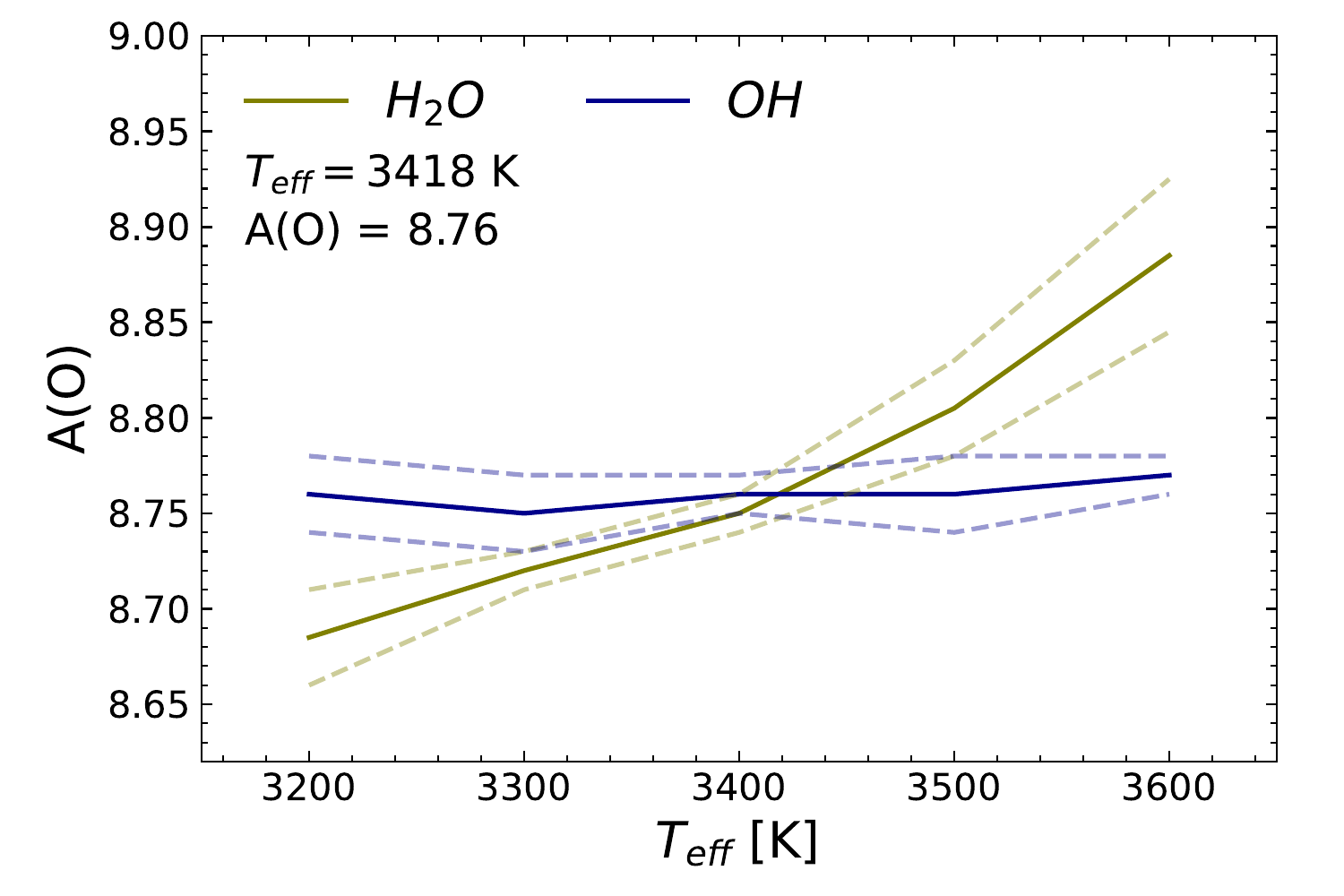}
\caption{An illustration of the dependence of the oxygen abundances derived from the OH and water lines as a function of the effective temperature of the model atmosphere. The two solid
lines show, respectively, the median values of the oxygen abundances derived from OH lines (blue) and H$_{2}$O lines (green). The intersection of these lines in the $T_{\rm eff}$ and A(O) diagram defines the effective temperature and the oxygen abundance for the star. Dashed lines represent the median absolute deviations of the oxygen abundances from the respective OH and H$_{2}$O lines.
}
\end{center}
\label{teffplot}
\end{figure}

Figure \ref{teffplot} illustrates this methodology, the green, and blue solid and dashed lines represent, respectively, the median and median absolute deviations (MAD) of the water and OH abundances obtained from line measurements; the crossing between the solid lines define the effective temperature. In Table \ref{Spectral_Lines} we present our measured OH spectral lines, along with their excitation potentials and $\log{gf}$s. Since the M dwarf near-infrared spectrum is severely blended with water, we are not able to assign a specific transition for our water measurements. We present in the Table, the regions used for the chi-square minimization used in the water abundances determination.

The metallicities and surface gravities for the studied stars were determined one at a time from best fits between model and observed spectra obtained via chi-squared minimization in 20 \AA\ wide spectral regions covering the entire APOGEE spectrum.
We note that our methodology to derive metallicities and surface gravities is different from \citet{souto2020} who used best fits for Fe I lines.
Projected rotational velocities (v$\sin{i}$) were also derived from individual chi-squared fits to individual OH lines.
Finally, we iterated the solution, by starting with the stellar atmospheres model obtained from the steps above and refining the solution by recomputing individual oxygen abundances for the OH and water lines. The derived spectroscopic parameters ($T_{\rm eff}$, $\log{g}$, [M/H], A(O) and v$\sin{i}$) are presented in Table \ref{stellartable}.

\begin{figure*}
\begin{center}
  \includegraphics[angle=0,width=1\linewidth,clip]{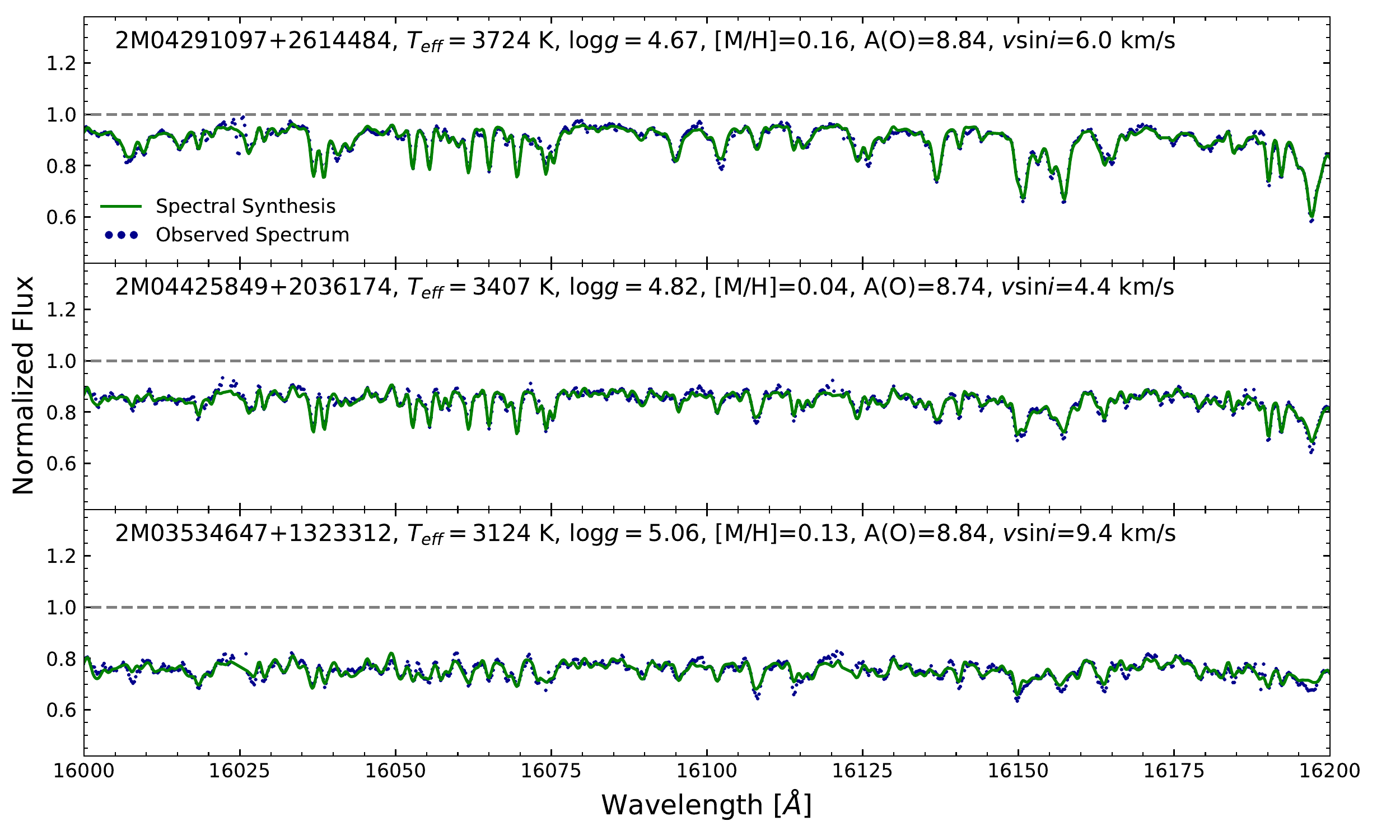}
\caption{Comparisons between observed APOGEE spectra and synthetic spectra spanning a 200\r{A} region within the APOGEE spectral window for the stars 2M04291097+2614484 (top panel), 2M04425849+2036174 (middle panel), and 2M03534647+1323312 (bottom panel). These spectra illustrate the changes in spectral features as the effective temperature decreases from spectral type M1 to M5. Note the deepening pseudo-continuum in the cooler M dwarfs, caused by increasing absorption from, primarily, water and FeH lines.}
\end{center}
\label{obssint}
\end{figure*}

An illustration of the quality of the model fits obtained in this study is presented in Figure \ref{obssint}, where we show the observed (blue points) and synthetic (green line) spectra for the APOGEE spectral region around 16,100 \AA\ obtained for three target M dwarfs covering the range in T$_{\rm eff}$ in this study: 2M04291097+2614484 (T$_{\rm eff}$= 3724 K; top panel); 
2M04425849+2036174 (T$_{\rm eff}$=3407 K; middle panel) and 2M03534647+1323312 (T$_{\rm eff}$=3124 K; bottom panel). 
Besides illustrating the APOGEE observed spectra of M dwarfs and the quality of the fits obtained as a function of the effective temperature, this figure also illustrates the deepening of the pseudo-continuum with decreasing T$_{\rm eff}$ in M dwarfs, which is due to increased absorption from water lines as well as iron hydride lines.

\subsection{Stellar Luminosities and Radii}

The methodology adopted to derive the stellar luminosities and stellar radii followed the previous study by \citet{souto2020} and as discussed in the latter, their derived radii were compared to and showed good agreement with stellar radii measured via interferometric measurements.
We obtained bolometric luminosities for our sample using K$_{\rm s}$-band bolometric corrections from \citet{mann2015,mann2016} photometric calibrations, V magnitudes from \citet{muirhead2018}, and Bailer-Jones EDR3 distances \citep{bailerjones2021} and an adopted zero point luminosity of $3.0128 \times 10^{35}$ erg s$^{-1}$ from \citet{mamajek2015}. 
The bolometric corrections and derived luminosities (in units of $10^{31}$ erg s$^{-1}$) are presented with the other stellar parameters in Table \ref{stellartable}.

Using the computed values for the stellar luminosities and the spectroscopic effective temperatures derived in this study, stellar radii were derived using the Stefan-Boltzmann equation, given below. 

\begin{equation}
    L=4 \pi R^{2} \sigma T_{\rm eff}^{4}
    \label{stefanboltzmann}    
\end{equation}

Our stellar radii derived from luminosity and T$_{\rm eff}$ are presented in Table 3 and these will be compared to theoretical radii from various isochrone families in the literature.

\subsection{Isochrone Families and Stellar Masses}

Available comparison isochrones (without magnetic fields) include those from MIST (MESA Isochrones and Stellar Tracks; \citealt{choi2016_MIST}) and DARTMOUTH \citep{Dotter2008_dartmouth}, along with PARSEC (V2.0; \citealt{Bressan2012_parsec,nguyen2022}) and BHAC15 \citep{baraffe2015}.  
As only solar-metallicity isochrones are presented in BHAC15, while the Hyades stars are significantly metal-rich, these particular isochrones are not used in the radii comparisons.  In addition, since the PARSEC isochrone radii have been adjusted to fit M-dwarf radii derived from eclipsing binary systems, while the MIST and DARTMOUTH isochrones are based entirely on stellar models, we will restrict our M-dwarf radii comparisons using MIST and DARTMOUTH.

An additional set of isochrones used for comparison with the derived M-dwarf radii is from \citet{Somers2020_spots} and are referred to as SPOTS models (Stellar Parameters Of Tracks with Starspots). These isochrones are somewhat different from those discussed above, as each SPOTS isochrone has an associated spot fraction, corresponding to different spot coverage, f$_{\rm spot}$, of the stellar photosphere.
The average temperature of the region covered by spots is 80\% of the temperature of the rest of the photosphere and the SPOTS effective temperatures are disk-integrated temperatures of the photosphere \citep{Somers2020_spots}.

Three stellar mass estimates were obtained by matching the observed luminosities to those from mass-luminosity values from the three families of isochrones discussed above (MIST, DARTMOUTH and SPOTS). Each mass estimate is presented in Table \ref{rotationaltable} along with average masses and standard deviations.

\subsection{Estimated Uncertainties}

In this section, we discuss different sources of uncertainties in the derived parameters for this study. The final uncertainties are: $\pm$ 100K for $T_{\rm eff}$; $\pm$ 0.15 dex for $\log{g}$; $\pm$ 0.14 dex for the derived metallicities;  $\pm$1.2 km s$^{-1}$ for v$ \sin{i}$, $\pm$0.10 for oxygen abundances and 0.005 M$_{\odot}$ for stellar masses. 

The uncertainties in the derived T$_{\rm eff}$s and oxygen abundances were estimated previously in \citet{souto2020} and their estimated uncertainties are adopted here.
The uncertainties in the derived $\log{g}$s and metallicities were estimated by summing in quadrature two uncertainty sources: the dispersion of the $\log{g}$ and metallicity values derived from the best-fits obtained for each individual spectral region analyzed (average uncertainty for $\log{g}$ and metallicity are respectively $\pm 0.09$ and $\pm 0.07$); and the uncertainty associated to our methodology's sensitivity to change in input parameters, uncertainties were obtained by perturbing each parameter individually and we obtained an average uncertainty for $\log{g}$ and metallicity of respectively $\pm 0.13$ dex and $\pm 0.12$ dex. 

For v$\sin{i}$ we considered the line-to-line dispersion of the OH molecular lines to compute the individual stellar uncertainties and adopted the mean of the individual results as the global uncertainty for the sample. We also computed the line-to-line dispersion for oxygen abundances, considering the analyzed OH and water lines. The line-to-line uncertainties for each star are presented in Table \ref{stellartable}.

In addition, we also investigated the effect of adding or subtracting the pixel-to-pixel flux uncertainties to the observations. However, this made a very small difference in the derived results, where we found typical uncertainties for T$_{\rm eff}$, $\log{g}$, [M/H], A(O) and v$\sin{i}$ of, respectively,  $\pm$14 K,  $\pm$0.03 dex,  $\pm$0.03 dex,  $\pm$0.04 dex and  $\pm$0.4 km s$^{-1}$, which are small and not representative of the expected uncertainties in our results.

Finally, we can infer internal uncertainties by investigating the scatter in the mean abundances of the Hyades cluster obtained from our sample.
The derived metallicities and oxygen abundances show good levels of consistency within the cluster stars; the standard deviations and MADs are 0.04 and 0.03, and 0.03 and 0.03 dex, respectively, for metallicities and oxygen abundances.

To compute the uncertainties in the radii, we used the uncertainties in distances from \citet{bailerjones2021} and the estimated uncertainties in the bolometric corrections of 0.036 mag, to obtain uncertainties in the stellar luminosities which, along with T$_{\rm eff}$ uncertainties, are propagated to obtain stellar radii uncertainties. These values vary between 0.01-0.03 $R_{\odot}$, and represent (median$\pm$MAD) 6.1$\pm$0.2\% of the stellar radii.

To estimate the uncertainties in the derived masses we used the median standard deviation of the median stellar mass obtained from the three sets of isochrones in Table \ref{rotationaltable} to compute a median uncertainty in mass for the sample of 0.005 M$_{\odot}$, which is equivalent to $\sim 1.5\%$ of the median stellar mass.

\section{Results}

\begin{figure}[h!]
\begin{center}
  \includegraphics[angle=0,width=1\linewidth,clip]{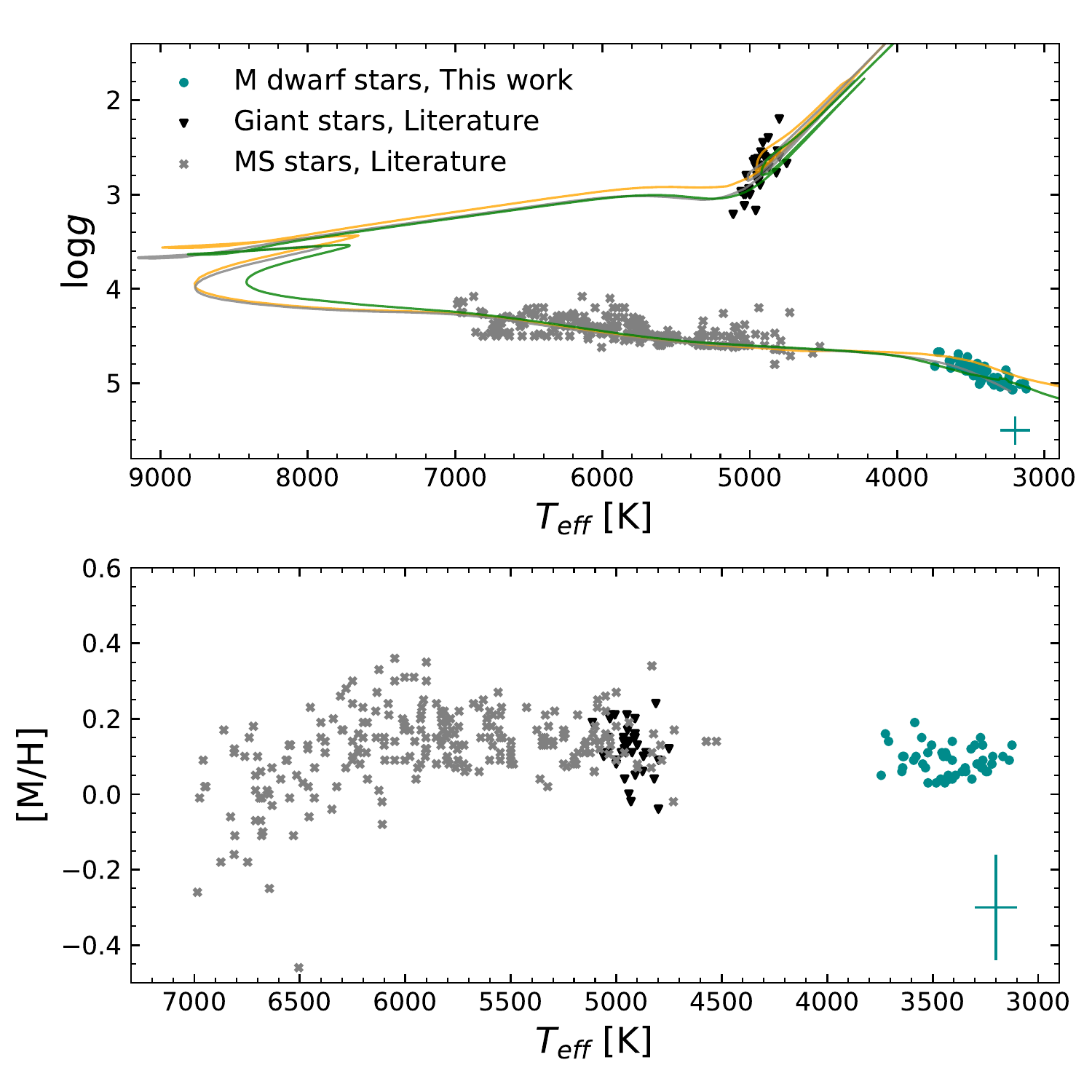}
\caption{Top Panel: A Kiel diagram showing the results for the Hyades M dwarfs obtained here from APOGEE spectra (cyan circles), along with those from high-resolution optical studies from the literature for Hyades FGK dwarfs (grey Xs) and red-giant stars (black triangles). Three sets of isochrones with an age of 0.6 Gyr are shown as guidelines: PARSEC (V2.0; yellow line, [M/H]=+0.07), DARTMOUTH (grey line, [M/H]=+0.07) and MIST (green line [M/H]=+0.1). Bottom panel: comparisons between the metallicities versus T$_{\rm eff}$ for the M dwarfs (this work), along with those from red giants and hotter dwarfs from the Hyades cluster, with these comparisons spanning almost 4000 K in T$_{\rm eff}$. The median metallicity for the M dwarfs studied here is $<[M/H]>=0.09\pm0.03$, and shows no significant trend as a function of T$_{\rm eff}$; it is also in good agreement with the median metallicity for the red giants from the literature ($<[M/H]>=0.12\pm0.03$).
The literature results reveal a dip in metallicity for the hotter dwarfs with effective temperatures between 6000 and 7000K. The estimated uncertainties are shown as errorbars.}
\end{center}
\label{metallicitylit}
\end{figure}

The studied M dwarfs cover roughly the spectral type range between M1 -- M5 (Section 2); all targets have effective temperatures higher than $T_{\rm eff} \sim 3100$ K, and lower than $T_{\rm eff} \sim 3800$ K, as above this effective temperature limit water lines start to become too weak for reliable measurements. 
The surface gravity values obtained from our analysis are typical of M dwarfs and range between roughly $4.7 \lesssim \log{g} \lesssim 5.1$.  
The $T_{\rm eff}$--$\log{g}$ results for the stars are presented in the Kiel diagram shown in the top panel of Figure \ref{metallicitylit} (shown as cyan circles). 
The figure also presents for illustration purposes only PARSEC (V2.0, \citealt{Bressan2012_parsec,nguyen2022}), DARTMOUTH \citep{Dotter2008_dartmouth} and MIST \citep{choi2016_MIST} isochrones corresponding roughly to the Hyades age (0.6 Myr), and with metallicities of respectively +0.07, +0.07, and +0.1.

The metallicity results for the studied Hyades M dwarfs are presented in the bottom panel of Figure \ref{metallicitylit} as a function of their effective temperatures.
The median ($\pm$ MAD) metallicity obtained for this sample is metal-rich relative to the Sun and the scatter in the metallicity measurements is small and can, in principle, be explained by the internal uncertainties in the determinations: $<$[M/H]$>$= +0.09$\pm$0.03; such consistency in the derived metallicities is in line with the expectation that stars in an open cluster are from a single stellar population and have homogeneous abundances. 

An important feature to highlight in our results shown in Figure \ref{metallicitylit} is the good consistency of the metallicities obtained here for a sample of M dwarf stars, which cover an extended range of $\sim$700 K in effective temperature. 

As discussed in the introduction, to our knowledge, this is the first high-resolution spectroscopic abundance study of M dwarf stars in the Hyades open cluster, but, there are several studies that analyzed the hotter main-sequence, and also red-giant stellar members of this cluster, and their metallicities can be compared to the M dwarf metallicity scale derived here.

Using the PASTEL database \citep{Soubiran2016_pastel} as a starting point and by adding results from the literature for Hyades members, we compiled a total of 286 metallicity measurements for stars in the Hyades open cluster (with $T_{\rm eff}$ in the range between 4500 -- 7000 K), that were analyzed in 37 published works from 1990 to the present (note that these are not all studies about the Hyades open cluster, but high-resolution studies that have results for at least one Hyades member. Also, we only considered stars with T$_{\rm eff}<7000$ K).
Figure \ref{metallicitylit} (top and bottom panels) shows results for the compiled sample, which are taken from the following studies: \citet{Boesgaard1990, Fernandez-Villacanas1990, McWilliam1990, Luck1995, King1996, Smith1999, Varenne1999, Cunha2000, Paulson2003, Sestito2003, Mishenina2004, Sestito2004, Valenti2005, Schuler2006, Randich2006, Mishenina2006, daSilva2006, Takeda2007, Hekker2007, Mishenina2008, Wang2009, Gebran2010, Kang2011, Pompeia2011, Carrera2011, Tabernero2012, Mishenina2012, Ramirez2013, Maldonado2013, Mortier2013, Ramirez2014, Datson2015, daSilva2015, Maldonado2016, Brewer2016, Aguilera-Gomez2018, Takeda2020}.

It can be seen from the top panel of Figure \ref{metallicitylit} that the literature sample is composed of stars on the main sequence that are cooler than the turnoff, as we selected stars with T$_{\rm eff}$ ranging roughly between 4500 -- 7000 K (shown as grey Xs), and more evolved Hyades members on the red-giant branch (shown as black triangles), with all of the literature results following, quite reasonably, the isochrone tracks for an age of 0.6 Gyr.

The bottom panel of Figure \ref{metallicitylit} shows
the same sample as in the top panel. The metallicity results for the M dwarfs in this study overall compare well with those from optical studies in the literature.
The median metallicity of the warmer FGK dwarfs from the literature is $<$[M/H]$>$ = $0.13\pm0.05$, which is just slightly more metal-rich than that we obtained for our sample of M dwarfs ($<$[M/H]$>$ = 0.0$\pm$0.03), but with a noticeably larger scatter among the warmer dwarfs, resulting from the generally decreasing trend of metallicity as T$_{\rm eff}$ increases from roughly 6000 to 7000 K.
If, for example, we compute median metallicities using a $T_{\rm eff}$ threshold at 6000 K, we have median metallicities of, respectively, $+0.1 \pm 0.08$ and $+0.14 \pm 0.04$ for effective temperatures above and below 6000 K. This trend with effective temperature may be due in part to systematic uncertainties but may be explained by mechanisms, such as atomic diffusion, which decreases the overall metallicity of the stars and is a function stellar mass and age. When the hydrogen core is exhausted, main-sequence stars evolve through the sub-giant and red-giant branches expanding their outer convective envelopes and erasing the chemical fingerprints produced by atomic diffusion. 
The evolved stars in Figure \ref{metallicitylit} are either clump giants or are ascending red-giants within a relatively small $T_{\rm eff}$ range.
The scatter in metallicity for these red giant results from the literature is smaller when compared to that of the dwarfs, having a median ($\pm$ MAD) metallicity of $+0.12 \pm 0.03$, which compares well with the median and MAD values for the M dwarfs, within the uncertainties. All in all, red-giant stars are good comparisons for the M dwarfs, as giants are not expected to exhibit decreased abundances resulting from atomic diffusion, with the only expected abundance changes being in C and N, due to first dredge-up \citep{lagarde2012,choi2016_MIST}. 

\section{Discussion}

Radius inflation is a term coined to describe a phenomenon that appears in a fraction of low-mass stars in which their derived radii are larger than predicted by physical models. 
Different works in the literature, employing different techniques to derive stellar radii and compare them with physical models, have found evidence for radius inflation. Here we briefly summarize some of them.

Eclipsing binary systems analyzed using both photometric light curves and spectroscopic radial-velocity curves provide opportunities to determine both the masses and physical radii of their stellar components. \citet{Garrido2019} analyzed light curves (but not radial-velocity curves) for 230 detached eclipsing binary systems and found a general trend of radius inflation for low-mass main-sequence stars, with a relative increase of inflation in lower stellar masses when compared to stellar evolutionary models. \citet{parsons2018} presented mass-radius results for 23 M dwarfs that were members of eclipsing binary systems with white dwarfs as the companion star; they note that the small size of white dwarfs provides very sharp eclipses, allowing for precise radii determinations of the M dwarfs. Relative to selected isochrones, these M dwarfs have measured radii that are 6.2\% larger than predicted, although this average has a large scatter of 4.8\% (one sigma). The inflation is observed in both partially convective stars, where the mean value is 4.0 $\pm$ 2.5\%, as well as in the cooler fully convective stars, with a mean inflation of 7.1 $\pm$ 5.1\%. \citet{parsons2018} also include 101 eclipsing binary M dwarf radii taken from several published sources in their summary discussion and obtain an average inflation value of 5 $\pm$ 5\% for the entire sample.  Examples of detailed analyses of individual eclipsing binary systems include \citet{TorresRibas2002} for YY Gem, \citet{LopezMoralesRibas2005} for GU Boo, \citet{han2017} for T-Cyg1-12664, or \citet{healy2019} for NSVS 07394765. YY Gem and GU Boo are both double M dwarf binary systems and the stars in YY Gem are found to be inflated by $\sim$20\%, while those in GU Boo are $\sim$10-15\% larger than predicted by models. The eclipsing binary T-Cyg1-12664 has one M dwarf as the secondary star and \citet{han2017} find it not to be measurably inflated with respect to models, within a few percent, while the double M-dwarf binary NSVS 07394765 has stars that fall within $\sim\pm$7\% of the chosen comparison isochrone \citep{healy2019}. Broadly speaking, the radii results for M dwarfs in eclipsing binary stars indicate a range of inflated radii from little to no inflation, up to values as large as about 20\%.

\citet{Kesseli2018} studied a sample of 88 rapidly rotating fully-convective M dwarfs from their R$\sin{i}$ distribution and found that while the hotter M dwarfs in their sample ($0.18 M_{\odot}<M<0.4 M_{\odot}$) had underestimated radii in the models by $\sim 6\%$, the M dwarfs with masses between $0.08 M_{\odot}<M<0.18 M_{\odot}$ had inflated radii by 13$\%$--18$\%$, indicating the existence of an offset trend between theoretical and observational measurements inside the fully-convective limit.

The young open clusters NGC 2264, NGC 2547, and NGC 2516 were studied in \citet{Jackson2016}. They obtained statistical radii from the R$\sin{i}$ distribution derived from v$\sin{i}$ and rotational period measurements and found radius inflation of up to $30\pm 10 \%$ for fully convective pre-main-sequence stars and $\sim10\%$ for ZAMS stars with radiative cores. 
\citet{Jackson2018,Jackson2019} expanded this work and obtained
average radius inflation of $\sim14\%$ for K and M dwarf stars in the Pleiades and $\sim6\%$ for M dwarf stars in the Praesepe open cluster. Also for the Pleiades, \citet{SomersStassun2017} employed SED fitting to derive stellar radii for solar-type stars and found radius inflation up to 20$\%$--30$\%$.

Previous studies in the literature have also noted correlations between the occurrence of radius inflation and magnetic activity indicators, such as H$\alpha$ emission, x-ray luminosity, rapid rotation, or the Rossby Number, $Ro$ (defined in equation \ref{eq3}). The latter can be used as an indicator for stellar activity and often used as a diagnostic for the efficiency of the dynamo mechanism (e.g.,  \citealt{reiners2012,jeffers2018,lanzafame2017,CaoPinsonneault2022}).
Stars with short rotational periods have not dissipated significantly their angular momentum through magnetized winds and are expected to have stronger magnetic fields, while cooler stars have longer convective turnover times, due to their deeper convective envelopes, which can sustain dynamos for longer timescales than hotter and more massive stars.

For the Hyades open cluster, in particular, evidence of radius inflation has been reported by \citet{jaehnig2019}. They computed stellar radii from photometry for a sample of 68 stars of F, G, K, and M spectral types and compared the derived radii with isochrones, finding average inflation factors for their fast-rotators sub-sample of $\sim 5\%$, with the most inflated member showing more than $\sim 20\%$ inflation. \citet{jaehnig2019} also found a stronger anti-correlation between radius inflation and Rossby Number than with rotational period, which was in agreement with the findings by \cite{noyes1984} that the Rossby Number correlates better with magnetic activity than with rotational periods alone.

\subsection{Radius Inflation in Hyades M dwarfs}

To infer the amount of radius inflation for the Hyades sample studied here, we follow the same prescription as \citet{jaehnig2019} and compute fractional differences between the derived radius and the isochrone radius: 

\begin{equation}
    R_{\rm frac}=\frac{R-R_{\rm iso}}{R_{\rm iso}}
    \label{eq2}    
\end{equation}

The isochrone radii were derived by matching the stellar luminosities with a reference isochrone corresponding to the age and metallicity expected for the cluster.

In addition, to gain insight into radius inflation in relation to activity and rotation period, we also employ the Rossby Number, $Ro$, which is defined as:

\begin{equation}
   Ro=\frac{P_{\rm rot}}{\tau_{\rm cz}}
    \label{eq3}    
\end{equation}

As discussed in Section 2, to avoid having binary stars in our sample as much as possible, we did not include stars having non-unique rotational period measurements, within a 20$\%$ threshold difference. However, not all stars in our sample had measured rotational periods; for 35 of them, P$_{rot}$ were available in:
\citet{stauffer2018,douglas2019,delorme2011_refdouglas4,hartman2011_refdouglas5,douglas2016_refdouglas7}. (We note that when more than one P$_{\rm rot}$ value was available an average was computed.)

To compute $\tau_{\rm tcz}$ we used Equation \ref{eqtcz}, which was obtained from an empirical calibration in  \citet{wright2018_tcz}, and used the stellar masses derived in Section 3.2 to compute convective turnover times, and finally Rossby Numbers:

\begin{equation}
    \log{\tau_{\rm cz}}=2.33-1.50 (M/M_{\odot})+0.31(M/M_{\odot})^{2} .
    \label{eqtcz}    
\end{equation} 

In Table \ref{rotationaltable} we present rotational periods, along with three values of fractional radius inflation, convective turnover times, and Rossby Numbers derived from different isochrones for the studied stars.

The sample was divided into three groups based upon rotational periods: ``slow-rotators'' having periods longer than 5 days (P$_{\rm rot}>$ 5 days, composed of 17 stars) and ``rapid-rotators'' with periods less than 5 days (P$_{\rm rot}<$ 5 days, composed of 18 stars), with the remaining stars in our sample having no rotational period information available in the literature (13 stars). If we separate the same sample based on Rossby Numbers instead of rotational periods, we would have a threshold of 0.06 with low Rossby Numbers representing high rotational velocity.

\begin{figure}[h!]
\begin{center}  \includegraphics[angle=0,width=1\linewidth,clip]{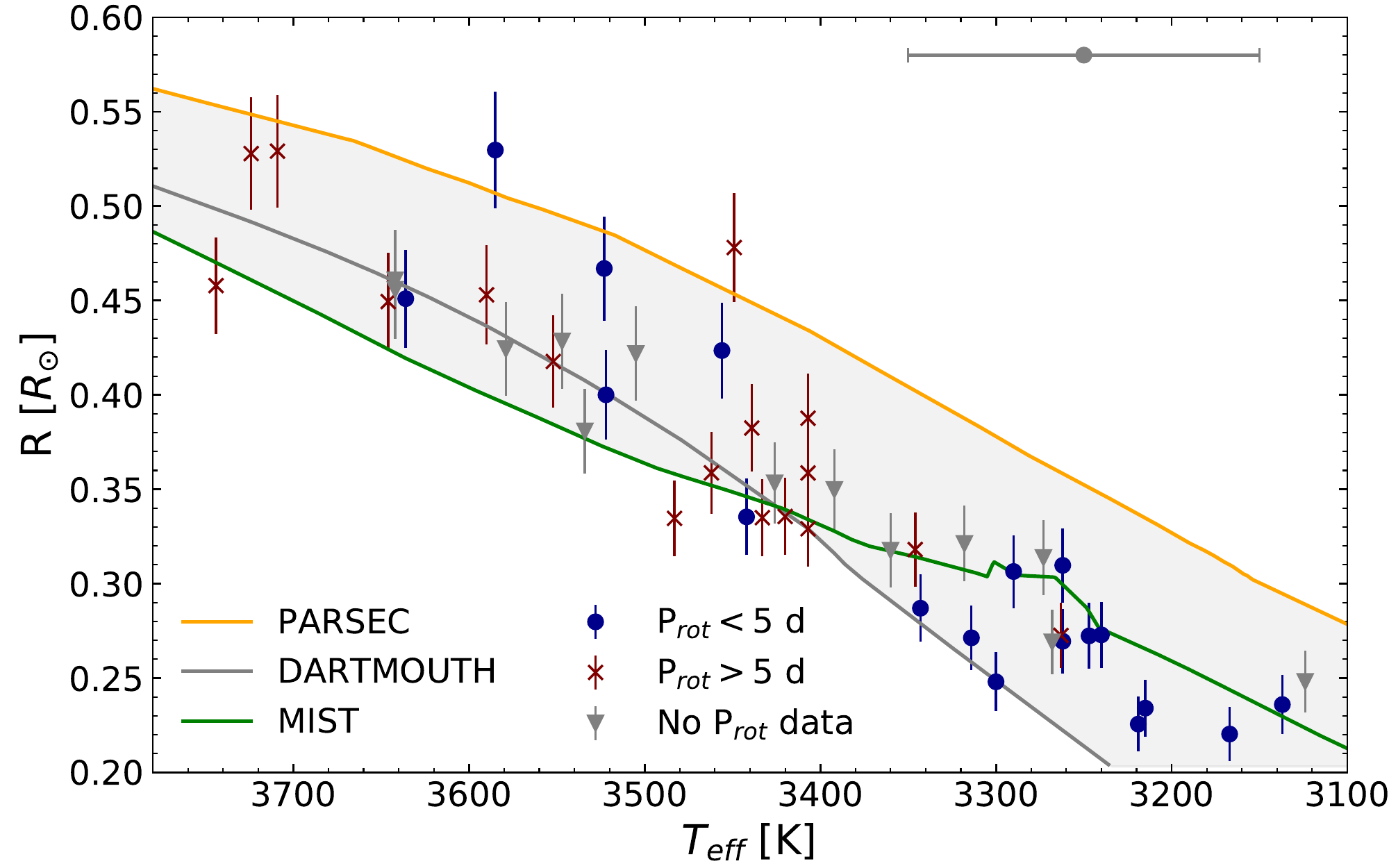}
\caption{Radii versus effective temperatures are shown for the M dwarfs segregated based on rotation (P$_{\rm rot}>5$ days are slow rotators and P$_{\rm rot}<5$ days are fast rotators) and mass (M$>$0.35 M$_{\odot}$ are partially convective M dwarfs and M$<$0.35 M$_{\odot}$ are fully-convective stars). The same set of three isochrones shown before (Figure \ref{metallicitylit}) are presented here. The shaded grey area represents the region covered by the isochrone limits, including the PARSEC one, which is the most discrepant. The estimated uncertainties are shown as errorbars.}
\end{center}
\label{radiusinflation_prot}
\end{figure}

Figure \ref{radiusinflation_prot} shows the stellar radius versus $T_{\rm eff}$ derived for the M dwarfs segregated by rotation, along with three sets of $\sim 0.6$ Gyr isochrones (without consideration of magnetic fields or spots, Section 3.3) from different groups: PARSEC (\citealt{Bressan2012_parsec,nguyen2022}; version 2.0; [M/H]=+0.07), DARTMOUTH (\citealt{Dotter2008_dartmouth}; [M/H]=+0.07), and MIST (\citealt{choi2016_MIST}; [M/H]=+0.1).
All stars in our sample are shown: fast rotators are represented by blue circles, slow rotators by maroon Xs, and stars without rotational period measurements by grey triangles (note that although rotational periods are not available for these stars, we measure their v$ \sin{i}$ and these values are presented in Table \ref{stellartable}). 

One general feature from the $T_{\rm eff}$ distribution for the fast- and slow-rotator groups studied here is that, overall, the fast-rotator sample tends to be cooler than the slow-rotator sample, which is expected due to the shorter spin-down timescales of the latter. 

Overall, a simple inspection of the distribution of the derived radii relative to the isochrones indicates that if, for example, the MIST or DARTMOUTH, isochrones were adopted as references, we would find that many of the Hyades M dwarfs in our sample show some level of radius inflation; however, if the PARSEC isochrone were adopted instead, the conclusion would be opposite, with most stars having smaller radii than predicted by the model. 
Most of the results for the M dwarfs obtained here fall within the region between the isochrones (shaded in grey) but this is mostly because of the larger radii predicted by the PARSEC isochrone. It is important to note, however, that this is because the PARSEC isochrones have been tuned to reproduce the mass-radius relation of eclipsing binaries \citep{Bressan2012_parsec,nguyen2022} and this produces larger radii.
In addition, it is clear that in the effective temperature regime between $\sim$ 3400 -- 3100 K, the T$_{\rm eff}$-radius behavior of the MIST isochrone overall matches most the results for the coolest M dwarfs in our sample. This will be further discussed in Section 5.1.1.

\begin{figure*}
\begin{center}
  \includegraphics[width=0.8\linewidth,clip]{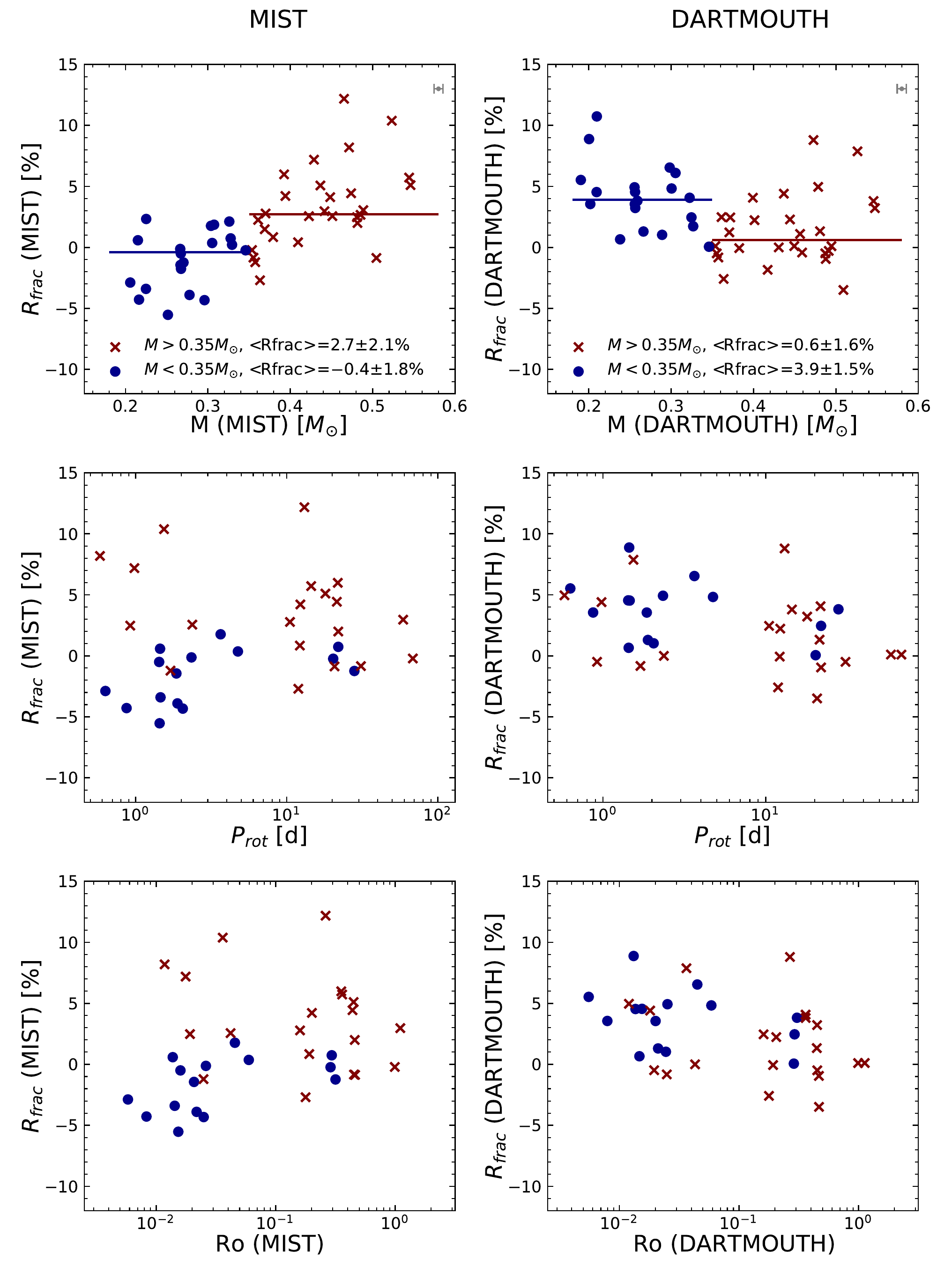}
\caption{Relation between the fraction of radius inflation relative to the MIST isochrone (left panels) and DARTMOUTH isochrone (right panels) as a function of mass inferred from the isochrone models (top panels), rotational period (middle panels) and Rossby Number (bottom panels) for the studied Hyades M dwarfs. The sample is divided into fully-convective M dwarfs with M $< 0.35$ M$_{\odot}$ (filled blue circles) and partially-convective M dwarfs with M $> 0.35$ M$_{\odot}$ (maroon Xs). The median ($\pm$  MAD) values of radius inflation are given for each sample and are represented by the horizontal lines in the upper panels. The estimated uncertainties are shown as errorbars.} 
\end{center}
\label{dartmistfig}
\end{figure*}

Figure \ref{dartmistfig} shows the fraction of radius inflation relative to both the MIST isochrone (left panels) and DARTMOUTH isochrone (right panels) as functions of mass (top panels), rotational period (middle panels), and Rossby Number (bottom panels) for the sample of Hyades M dwarfs. In this figure, the sample is divided by mass into fully-convective M dwarfs, with M $<$ 0.35 $M_{\odot}$ (filled blue circles), and partially-convective M dwarfs having radiative cores and convective outer envelopes (M $>$ 0.35 $M_{\odot}$; maroon Xs).
When considering the results for the entire Hyades M dwarf sample, the MIST, and DARTMOUTH isochrones indicate a median ($\pm$ MAD) radius inflation of respectively 1.6$\pm$2.3\% and 2.4$\pm$2.3\%, and we find radius inflation values of up to, respectively, $\sim$12\% and $\sim$11\%. However, these statistics that overall indicate good agreement do not highlight important differences that depend on the model. In the following subsections, we will first discuss the comparisons with the MIST isochrone (Section 5.1.1), and subsequently the results relative to the DARTMOUTH isochrone (Section 5.1.2), and finally relative to the SPOTS models (Section 5.1.3).

\subsubsection{Radius Inflation relative to MIST models}

Focusing solely on the left panels of Figure \ref{dartmistfig} showing fractional radius inflation ($R_{\rm frac}$) when adopting the MIST isochrone as a baseline, the median radius inflation obtained for the entire sample of 48 Hyades M dwarfs is 1.6\%. However, the segregation by mass in this figure reveals a distinction between the behavior of the fully convective M dwarfs relative to the partially convective ones. Stars with M$<$0.35 $M_{\odot}$ show basically no radius inflation (a median radius inflation of -0.4\%$\pm$1.8\%), while the more massive, partially convective M dwarfs (M $>$ 0.35 $M_{\odot}$) have a median value of 2.7 $\pm$2.1\% for the radius inflation. This point is illustrated in the top panel of Figure \ref{dartmistfig}, where we show horizontal lines in blue (for M$<$0.35 $M_{\odot}$) and maroon (for M$>$0.35 $M_{\odot}$) that represent the median values of fractional radius inflation for each case.

The fraction of radius inflation as a function of rotational period and Rossby Number is shown in the middle and bottom panels of Figure \ref{dartmistfig}, respectively. The statistics based on rotational periods show similar behavior as above, with the rapid- (P$_{\rm rot}<$ 5 days) and slow-rotator (P$_{\rm rot}>$ 5 days) samples having median radius inflation values of -0.3$\pm$2.8\% and 2.0$\pm$2.4\%, respectively. For the M dwarfs with radiative cores (represented by maroon Xs) in particular, we can see a trend where the six stars with P$_{\rm rot}<$ 5 days (or $Ro\lesssim$ 0.06) present a median inflation of 4.9$\pm$2.9\%, while the slow-rotators present a median inflation of 2.9$\pm$2.5\%. Given the small number of statistics, this difference may not be considered significant for our sample.
However, we compared our results to those from \citet{jaehnig2019} by deriving stellar masses for their sample using their published bolometric fluxes to derive luminosities, along with their published effective temperatures, and found that their entire sample has estimated masses $M>0.35 M_{\odot}$. Our results for partially-convective M dwarfs are then in agreement with their results, with both having similar dependences between radius inflation and rotational periods and Rossby Numbers. This correlation between radius inflation and rotational period for Hyades M dwarfs with radiative cores suggests that radius inflation, in these stars, is possibly correlated to activity and magnetic fields. 

The absence of radius inflation for fully convective M dwarfs in the left panels of Figure \ref{dartmistfig}, even though these stars have low Rossby Numbers (middle panel of Figure  \ref{dartmistfig}), is in line with predictions of stellar models that additionally include magneto-convection in the stellar interior structure.
\citet{feiden2015} presented Dartmouth magnetic isochrone models that include the effects of magnetic fields. They showed that either rotational or turbulent dynamo prescriptions predict that the radii of fully convective stars are almost unaffected by magnetic fields, with only very strong internal magnetic fields being capable of inflating these stars. On the other hand, magnetic inhibition of convection can produce significant inflation for stars with radiative cores and convective outer envelopes \citep{feiden2012,feiden2013,macdonald2014}. As discussed in \citet{chabrier2007}, magnetic fields can inflate a star by inhibition of convection or by producing large stellar spots that can inflate a star due to its unchanged total luminosity. However, the radii of fully-convective M dwarfs would only be affected by the latter effect, and only if a large percentage of the stellar photosphere is covered by spots (around 30--50 $\%$).
 
Finally, we point out that, although the median value for $R_{\rm frac-MIST}$ is near zero for the fully convective stars in our sample, there may be a hint of a possible mass dependency for radius inflation inhibition within the fully convective regime. 
It should be kept in mind that the boundary for a star becoming fully convective may be uncertain and this combined with systematic uncertainties in the masses may make the classification of stars near the boundary incorrect.

As shown in Figure \ref{dartmistfig}, there is a similar relation between radius inflation with rotational periods and Rossby Numbers. This relationship was also studied in other works, \citet{jaehnig2019} binned their Hyades results in rapid- and slow-rotators considering a threshold in Rossby Numbers of 0.1, and found that it produces a better separation between slow- and fast-rotators than when considering rotational periods. They also studied the results from \citep{SomersStassun2017} for solar-type Pleiades stars and found the same threshold in Rossby Numbers. As previously discussed, we found a transition at P$_{\rm rot} \sim 5$ days, or $Ro\sim0.06$. We note, however, that there is a gap in our sample between slow- and fast-rotators, with the fastest star from the slow-rotator sample presenting a rotational period of 10.5 days and a Rossby Number of 0.16.  

\subsubsection{Radius Inflation relative to DARTMOUTH models}

Similarly to what was presented and discussed in the previous section using the MIST isochrone as a reference, the right panels of Figure \ref{dartmistfig} now show $R_{\rm frac}$ results pertaining to comparisons with a DARTMOUTH isochrone. 
An overall median fractional radius inflation value of $2.4\%$ is obtained when considering the full sample of Hyades M dwarfs and this median $R_{\rm frac-DARTMOUTH}$ value is similar to that found for the full sample relative to the MIST isochrone.  This simple comparison masks significant differences between the MIST and DARTMOUTH derived values of R$_{\rm frac}$ as a function of stellar mass, however, as is shown clearly in the top panels of Figure \ref{dartmistfig}. In the top right panel of Figure \ref{dartmistfig} (DARTMOUTH), the stars displaying significant non-zero radius inflation are the fully convective M dwarfs (with the median value of the inflation indicated by the blue horizontal line), as opposed to much smaller, or insignificant values of radius inflation for the partially convective M dwarfs (median value shown as the horizontal red line); this is exactly opposite to the behavior of  R$_{\rm frac}$ as a function of stellar mass derived from the MIST isochrone (top left panel), where the fully convective M dwarfs, on average, show no overall radius inflation.  In addition to the differences in R$_{\rm frac}$ resulting from comparisons between the baseline MIST and DARTMOUTH isochrones, we point out that \citet{feiden2015} described an update to the Dartmouth stellar evolution models that include magnetic fields generated by either a stellar rotational or turbulent dynamo. The results of this upgrade are summarized in Figure 2 of their work, where it can be seen that, relative to a baseline Dartmouth isochrone with an age of 1.0 Gyr and solar metallicity, radius inflation at levels up to $\sim$8\% (for a rotational dynamo) and $\sim$13\% (for a turbulent dynamo) can be generated, depending on the strength of the magnetic fields and stellar mass. Although the values of radius inflation increase with stellar mass for the partially-convective M dwarfs with radiative cores, the upgraded ``magnetic'' Dartmouth model does not predict significant radius inflation for the fully convective low-mass stars. 
However, although not predicted by the models, finding higher inflation for the coolest stars is in general agreement with many observational results from the literature (e.g., \citealt{parsons2018,Kesseli2018,Garrido2019}).

Values of R$_{\rm frac}$ derived from the DARTMOUTH isochrone versus $P_{\rm rot}$ and $Ro$ are shown in the right middle and bottom panels of Figure \ref{dartmistfig}. Using the same separation in $P_{\rm rot}$ at 5 days (or $Ro \sim $0.06) as previously, the median $R_{\rm frac-DARTMOUTH}$ for rapid- and slow-rotators are, respectively, 4.5$\pm$1.6\% and 1.3$\pm$1.8\%, indicating that stars that rotate fast have, in the median, more radius inflation. A similar result is found when considering the Rossby Numbers. Finally, we note that there is no clear distinction between the behavior for the fully and partially convective samples.

Before discussing the SPOTS models (Section 5.1.3) we briefly summarize the differences between the analysis from MIST and DARTMOUTH isochrones. By obtaining R$_{\rm frac}$ from different isochrones with the same ages (0.6 Gyr) and similar metallicities (0.1 for MIST and 0.07 for DARTMOUTH), we concluded that inflation is dependent on the prescriptions/constraints adopted for the different models. By dividing the data into fully- and partially-convective M dwarfs, we discovered that the outcomes from both isochrones concur that for partially-convective stars, radius inflation is strongly correlated with rotation. Stars with smaller Rossby Numbers and shorter rotational periods exhibit higher levels of inflation. The results were found to differ when it comes to fully-convective stars. The DARTMOUTH results suggest a correlation between rotation and inflation similar to that found for partially-convective stars, while the MIST isochrones predict well the radii for fully convective stars, implying that despite their fast rotation, fully-convective stars do not exhibit significant inflation levels, which is in agreement with models that include magnetic fields from \citet{feiden2015}.

\subsubsection{Radius Inflation relative to SPOTS models}

\begin{figure}[h!]
\begin{center}  \includegraphics[angle=0,width=1\linewidth,clip]{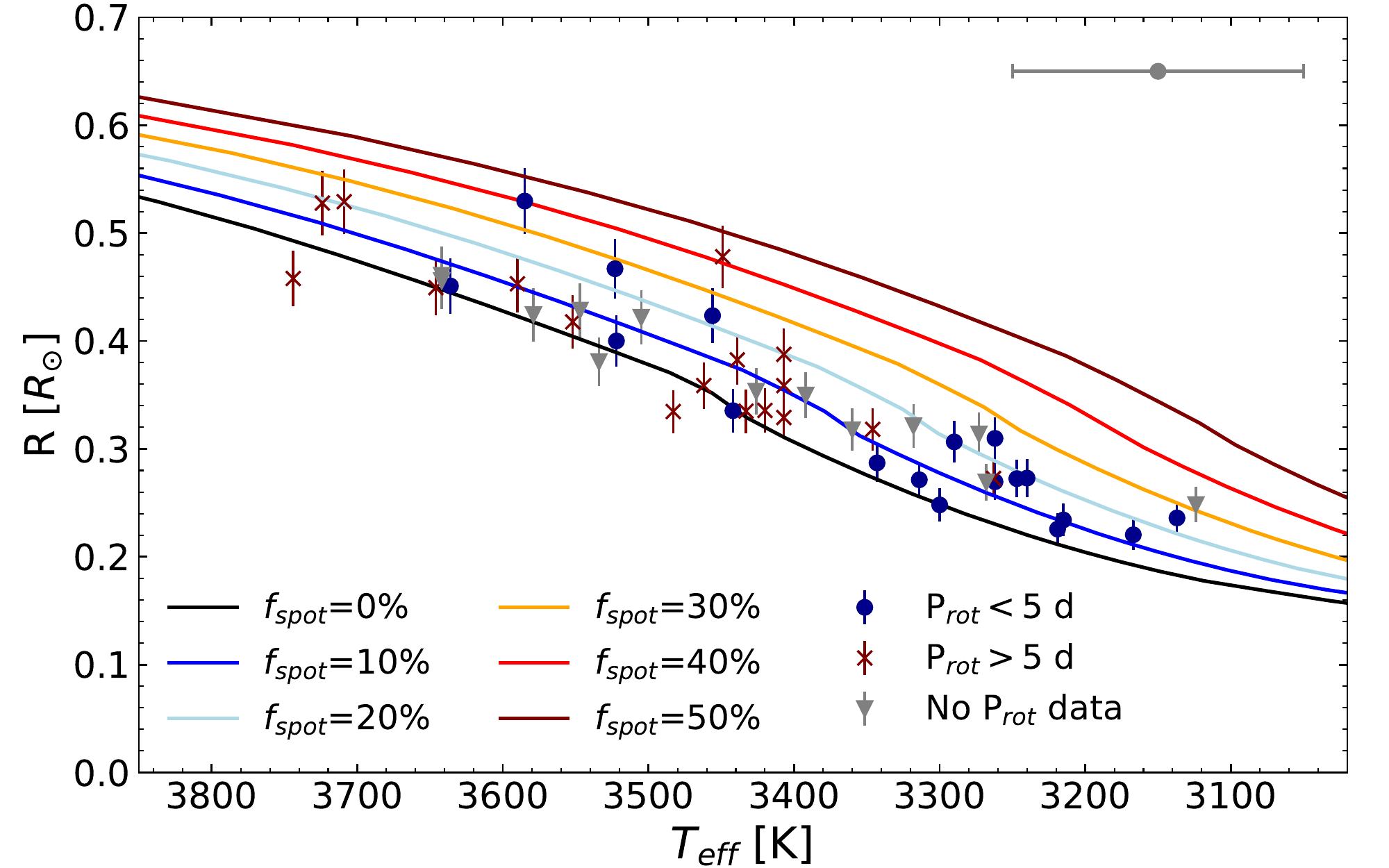}
\caption{Radii versus effective temperatures are shown for the M dwarfs segregated based on rotation (P$_{\rm rot}$=5 d separating slow and fast rotators). We also present SPOTS isochrones ([M/H]=0.1, Age$\sim$0.6 Gyr), with each curve representing stellar photospheres covered by varying spot fractions (f$_{\rm spot}$). The range in radius inflation found within our results can be explained generally in terms of models taking spots into account, with 76$\%$ of the studied M dwarfs having up to 20$\%$ spot coverage, and extremely inflated stars with 20 to $\sim$40$\%$ spot coverage. The estimated uncertainties are shown as errorbars.}
\end{center}
\label{radiusinflation_prot_spots}
\end{figure}

An additional comparison of the derived M dwarf radii with a different set of models is presented in Figure \ref{radiusinflation_prot_spots}. This figure shows [M/H]=+0.1 isochrones from \citet{Somers2020_spots} (SPOTS, for Stellar Parameters Of Tracks with Starspots) for an age of $\sim$ 0.6 Gyr; the SPOTS models were discussed in Section 3.4. Each isochrone in Figure \ref{radiusinflation_prot_spots} is associated with a different fraction of a stellar photosphere covered by spots (f$_{\rm spot}$, indicated by the percentage described in the legend of the plot).
Keep in mind that \citet{Somers2020_spots} used an average temperature of the region covered by spots to be 80\% of the temperature of the rest of the photosphere, with the SPOTS effective temperatures being disk-integrated temperatures of the photosphere.

As discussed previously, there is a class of models that produce larger radii from the inclusion of magnetic fields \citep{feiden2013,feiden2015}.
While \citet{feiden2015} accounted for magnetic fields in their models, the SPOTS models from \citet{Somers2020_spots} consider the effect of the spots on the structure of the star. The SPOTS models result in radius inflation by suppressing the convection in sub-surface layers and thus changing the temperature and pressure conditions in the photosphere. Differently from \citet{feiden2015}, which considers that luminosity is unchanged as stars inflate, 
SPOTS models result in stellar spots reducing the internal pressure within the star which, in the case of fully-convective stars, reaches the nuclear burning core, diminishing the fusion reactions, and therefore reducing the luminosity of the star \citep{SomersPinsonneault2015}. 

Figure \ref{radiusinflation_prot_spots} shows stellar radius versus effective temperature from SPOTS models, along with those derived for the M dwarfs and illustrates that, at a given effective temperature, SPOTS models with higher stellar spot fractions are associated with larger stellar radii. All Hyades M dwarfs studied here span ranges in spot coverage up to $\sim$40\%, although 76\% fall below the 20\% coverage models. This comparison demonstrates that stellar spot models can also explain the observed radius inflation in the Hyades M dwarfs.  

\begin{figure*}
\begin{center}  \includegraphics[angle=0,width=0.95\linewidth,clip]{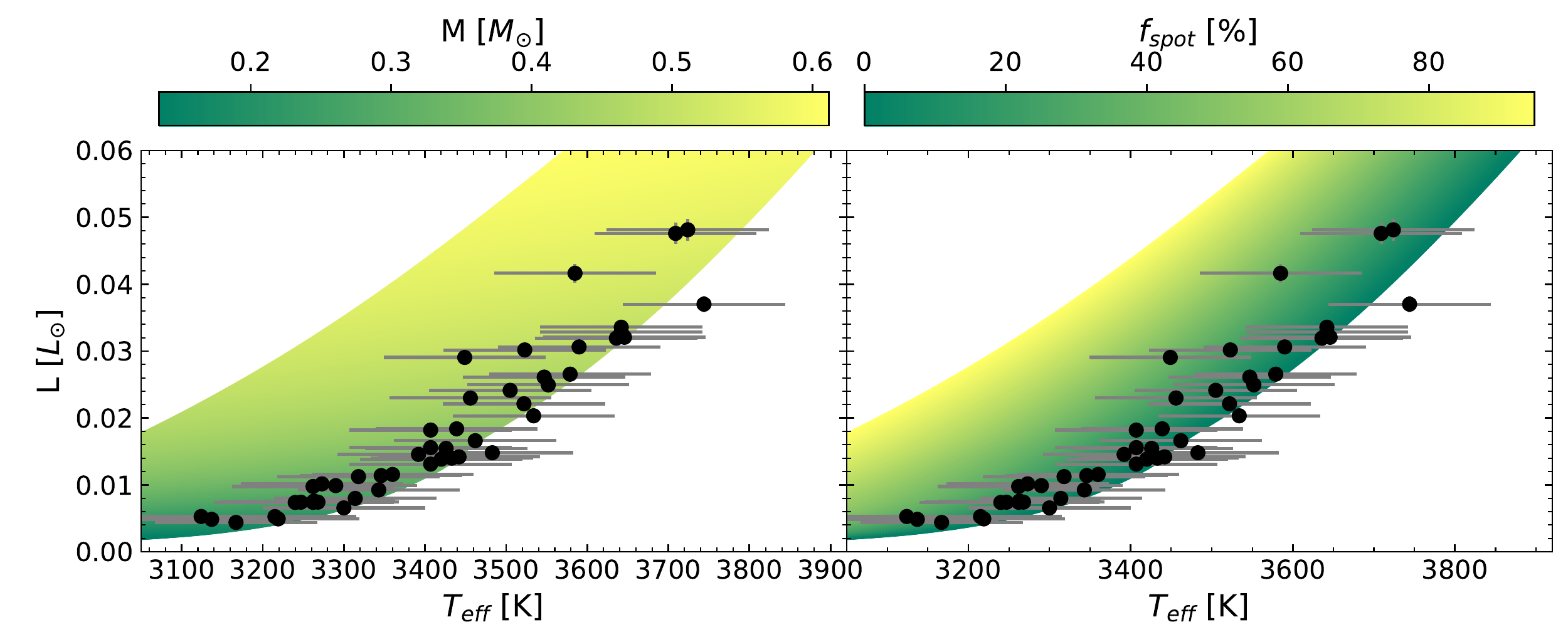}
\caption{Comparison between our derived effective temperatures and luminosities and SPOTS isochrones (age of $\sim$0.6 Gyr and metallicity of 0.1) interpolated in T$_{\rm eff}$ and luminosity. The isochrones in the left panel are color-coded by stellar mass, while in the right panel, they are color-coded by f$_{\rm spot}$. The adopted stellar mass and f$_{\rm spot}$ for each star is indicated by the location in the T$_{\rm eff}$--Luminosity plane. The estimated uncertainties are shown as errorbars.}
\end{center}
\label{spotscolorbar}
\end{figure*}

SPOTS isochrones predict that stars with higher f$_{\rm spot}$ values present lower luminosities if compared to less-spotted stars with the same mass. To account for this effect, instead of directly comparing stellar luminosities to the isochrones 
as previously, we interpolated SPOTS isochrones with different fractional spots coverage, in T$_{\rm eff}$ and luminosity, creating an isochrone plane. This is illustrated in Figure \ref{spotscolorbar}, showing the dependence of the effective temperatures with luminosities for SPOTS isochrones as functions of mass (left panel) and f$_{\rm spot}$ (right panel). We also show in Figure \ref{spotscolorbar} the effective temperatures and luminosities (and estimated uncertainties) for our stars as black filled circles.
Given a T$_{\rm eff}$-L pair the left panel of Figure \ref{spotscolorbar} was used to interpolate the mass, while the right panel was used to obtain f$_{\rm spot}$.
Finally, with this given mass, we considered R$_{\rm iso}$, as the radius for the same mass in the fiducial isochrone having zero spots (f$_{\rm spot}$ = 0\%), and compared with our stellar radius to derive R$_{\rm frac}$-SPOTS. (We note that three outliers with T$_{\rm eff}$-L out of the isochrones boundaries, which would require extrapolations, were excluded from subsequent analysis).

\begin{figure}
\begin{center}  \includegraphics[angle=0,width=1\linewidth,clip]{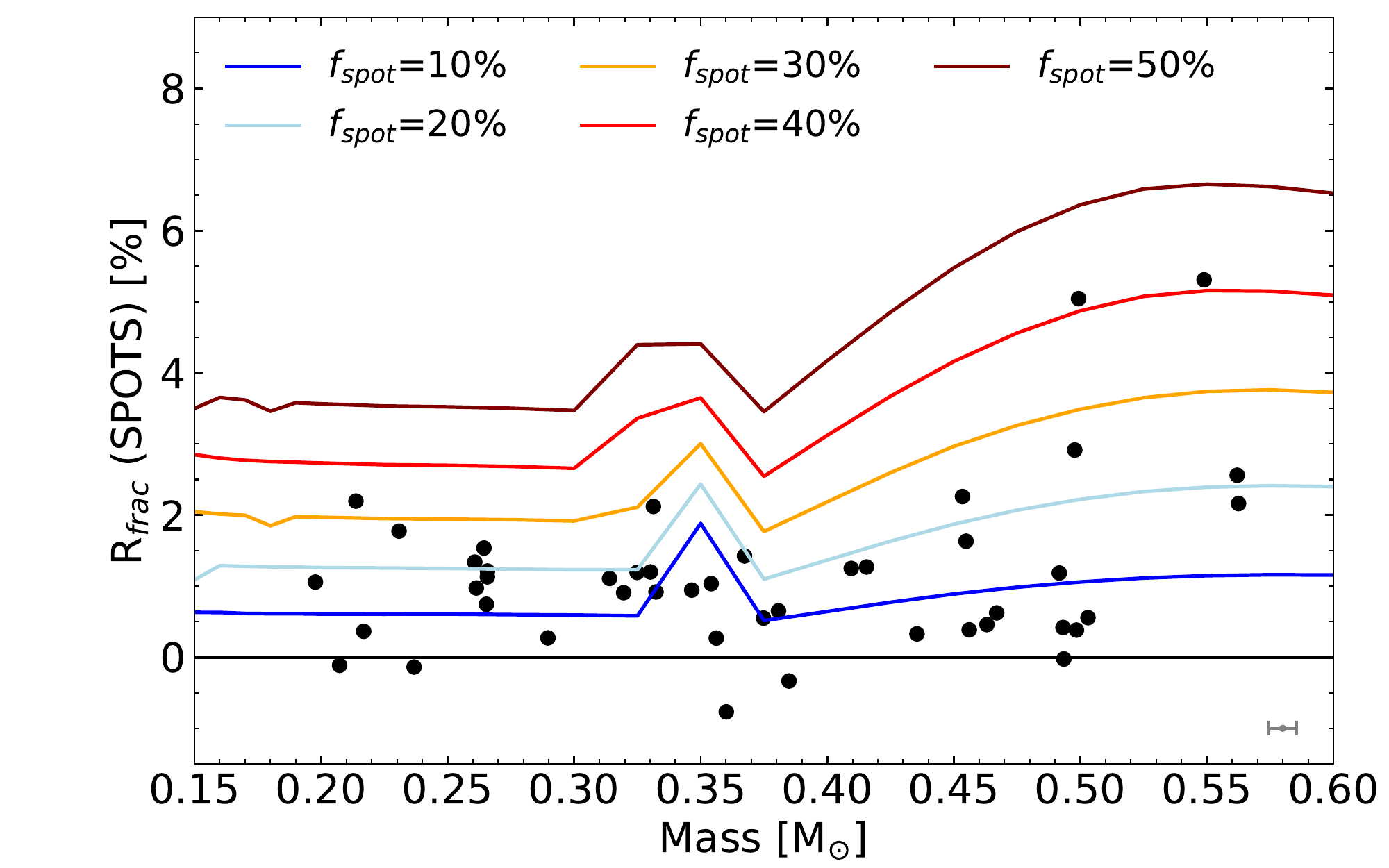}
\caption {Radius inflation relative to SPOTS isochrones (age=0.63 Gyr and [M/H]=0.1) versus masses are shown. Additionally, we show the radius inflation of SPOTS isochrones with varying levels of spot coverage (f$_{\rm spot}$) compared to a SPOTS isochrone with $0\%$ spot coverage. The estimated uncertainties are shown as errorbars.}
\end{center}
\label{rfracsp}
\end{figure}

Figure \ref{rfracsp} presents the distribution of masses and R$_{\rm frac-SPOTS}$ for our sample along with the isochrones R$_{\rm frac-SPOTS}$, obtained by comparison of the radii and masses of each isochrone and the f$_{\rm spot}=0 \%$ isochrones. The radius inflation obtained from SPOTS isochrones is considerably smaller than the ones obtained from MIST and DARTMOUTH isochrones, presenting a median ($\pm$  MAD) of 1.0 $\pm$ 0.5\%, with the most inflated star reaching an inflation value of 5.3\%.
The median inflation of the fully- and partially-convective samples are 1.1 $\pm$ 0.2\% and 0.7 $\pm$ 0.6\%, respectively. Although this single statistic for both sub-samples is very similar, it hides important differences, which can be seen in Figure \ref{rfracsp}.
Overall, fully-convective stars exhibit moderate levels of inflation ($\sim$0--2\%), despite most of them being rapid rotators with a small scatter, and inflation levels showing no clear dependence on stellar mass.
On the other hand, partially-convective M dwarfs exhibit significant scatter that is dependent on stellar mass, with the more massive examples reaching inflation levels of more than 5\%. The boundary between the fully convective and partially convective stellar models (M$\sim$0.34--0.37M$_{\odot}$) exhibits a small ``bump'' in R$_{\rm frac-SPOTS}$ which may be related to the non-equilibrium $^{3}$He-burning instability found by \citet{vansaders2012}.  

Our results are also in good agreement with what is predicted by SPOTS isochrones. A singular SPOTS isochrone with a given f$_{\rm spot}$ predicts an approximately flat relation between inflation and mass for fully-convective stars and a positive correlation for partially-convective stars. It also predicts a ``bump'' around the boundary between fully- and partially-convective M dwarfs. This behavior is similar to what is shown in our derived radii. 

\subsection{UV and X-Ray Activity in the Hyades Open Cluster}

Active M-dwarf stars have high-energy luminosities that are often orders of magnitude greater than what is expected from their stellar effective temperatures. 
Here we investigate whether the stellar luminosities in the near-ultraviolet (NUV), far-ultraviolet (FUV), and X-ray correlate with the derived spot fractions.
To compute high-energy luminosities for our targets we used the compiled X-ray luminosities in \citet{Freund2020_xrays}, and NUV and FUV data from the \textit{Galaxy Evolution Explorer} (GALEX) in \citet{Schneider2018_uv}. The computed high-energy luminosities (in units of $10^{27}$ erg s$^{-1}$), when available, are presented in Table \ref{rotationaltable}.

\begin{figure}[h!]
\begin{center}
  \includegraphics[angle=0,width=1\linewidth,clip]{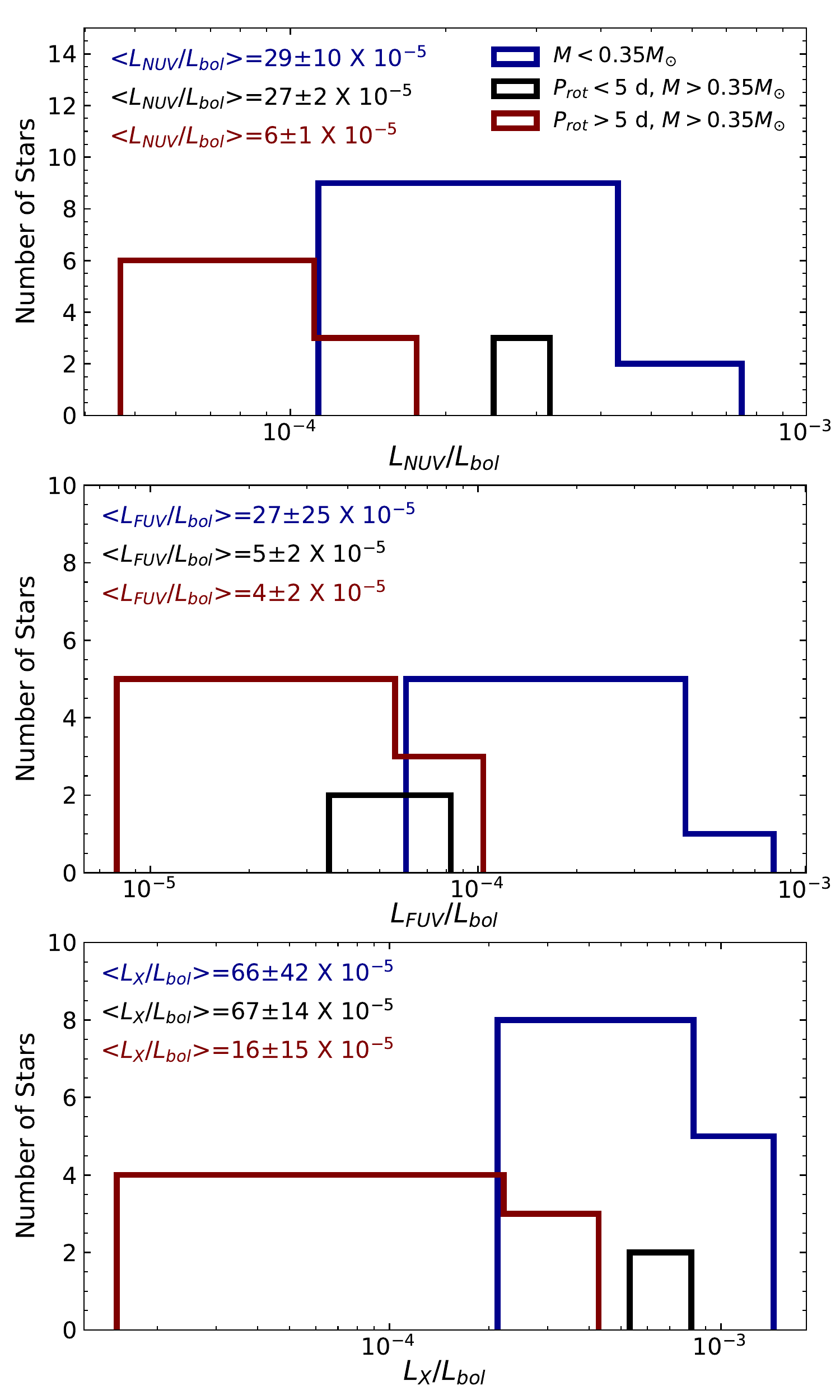}
\caption{Histograms of the ratios between the near-ultraviolet (top panel), far-ultraviolet (middle panel), and x-rays (bottom panel) luminosities and bolometric luminosities. The partially-convective rapid-rotator ($P_{\rm rot} < 5$ days), and slow-rotator ($P_{\rm rot} > 5$ days) samples are shown respectively in blue and green, and the fully-convective sample, in red. The median $\pm$ MAD $L_{\nu}$/$L_{bol}$ ratio of each dataset is given.}
\end{center}
\label{activity}
\end{figure}

Figure \ref{activity} illustrates the distribution of the NUV (top panel), FUV (middle panel), and X-ray (bottom panel) luminosities divided by the bolometric luminosities for partially-convective rapid rotators (black histogram), partially-convective slow rotators (red histogram), and fully-convective stars (blue histogram).  The median $\pm$ MAD of the $L_{\nu}/L_{bol}$ ratios are also given for each sample.
The partially-convective slow-rotators have typical luminosity ratios that are several times smaller than the other two samples, indicating that for our studied stars, activity is highly dependent on stellar rotation. The luminosity ratios of the rapidly-rotating partially-convective and fully-convective M dwarfs (the latter composed mostly of rapid-rotators) are very similar, with the exception of the FUV regime, where the fully-convective stars have higher ratios.

\begin{figure}[h!]
\begin{center}
  \includegraphics[angle=0,width=1\linewidth,clip]{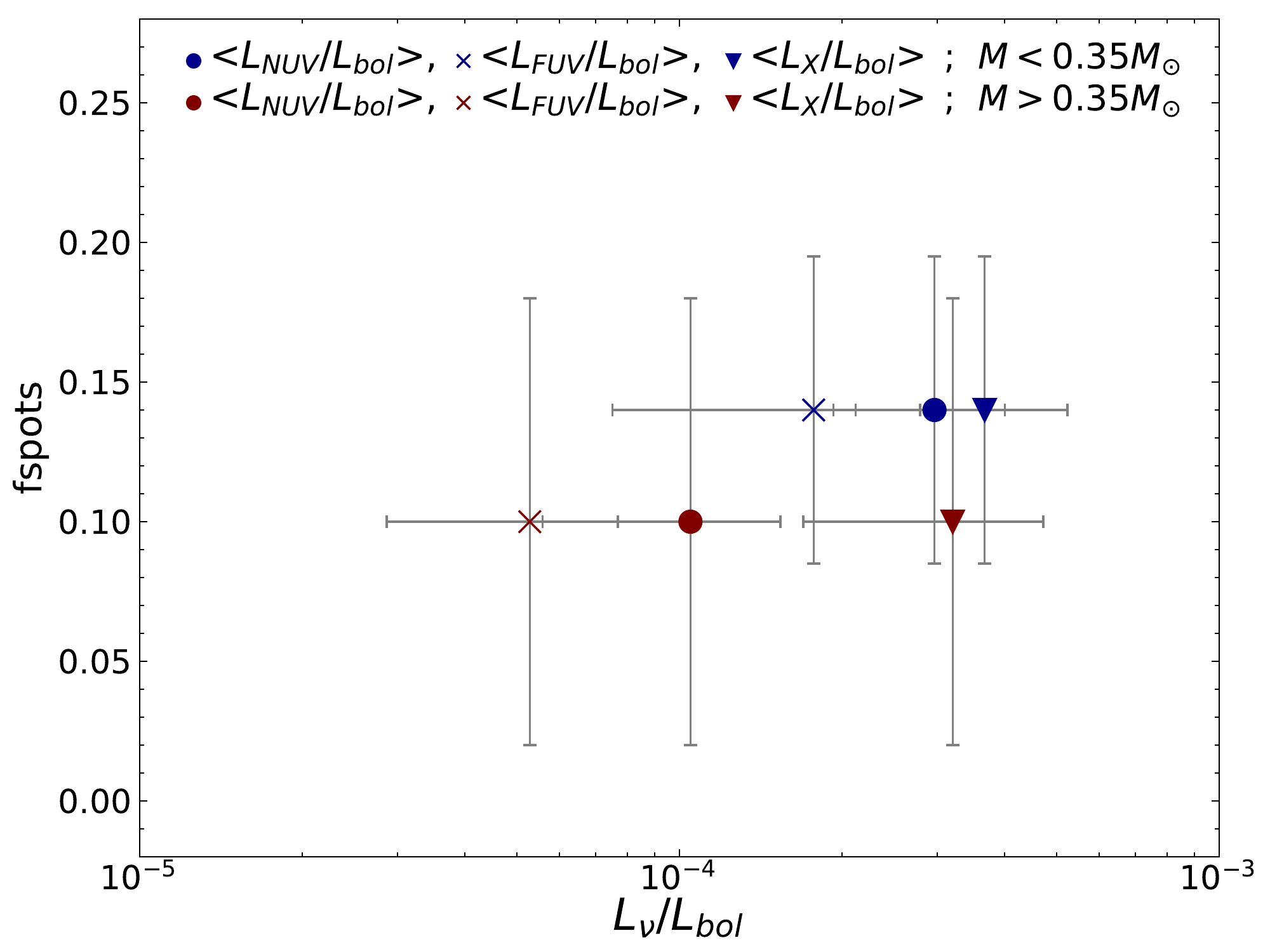}
\caption{Fractional spot coverage, f$_{\rm spot}$, versus the ratios of NUV, FUV and X-ray to bolometric luminosities for the Hyades M dwarfs segregated by mass into partially-convective and fully-convective stars.  The filled symbols are the median$\pm$MAD values for the fully- and partially-convective stellar samples (respectively in blue and maroon), and are also separated into the three high-energy luminosity regimes (NUV, FUV and X-ray) represented, respectively, by circles, crosses, and triangles).The estimated uncertainties are shown as errorbars.}
\end{center}
\label{activityxfspots}
\end{figure}

A comparison of the M dwarf surface spot fractions derived in this study with stellar activity levels is shown in Figure \ref{activityxfspots}, where f$_{\rm spot}$ is plotted versus the high-energy-to-bolometric luminosity ratios, L$_{\nu}$/L$_{\rm bol}$, with the sample divided into two groups: partially convective (M $<$ 0.35 M$_{\odot}$) and fully convective, (M $>$ 0.35 M$_{\odot}$) stars, using masses derived from the SPOTS isochrones.  The symbols and error bars in the figure represent median$\pm$MAD values of f$_{\rm spot}$ and L$_{\nu}$/L$_{\rm bol}$ for each mass bin and show that the fully-convective sample exhibits both higher activity levels and stellar surface spot fractions relative to the partially-convective sample. 
Note that the differences between the luminosity ratios in the two samples are smaller for the X-ray window when compared to the NUV and FUV differences, although all show a positive correlation between f$_{\rm spot}$ and stellar activity.

\section{Conclusions}

We performed a quantitative spectroscopic analysis on a sample of 48 single M-dwarf members of the young Hyades open cluster using SDSS/APOGEE high-resolution near-infrared spectra \citep{majewski2017_apogee} and present derived effective temperatures, surface gravities, metallicities, and projected rotational velocities. The analysis was based on the spectroscopic techniques, which rely on measurements of H$_{2}$O and OH lines, developed by \cite{souto2020,souto2022}. 

The Hyades M dwarfs analyzed here span the effective temperature range from T$_{\rm eff}\sim$3100 to 3750 K and have metallicities that compare well with previous literature determinations from high-resolution optical spectra that were based on either hotter main-sequence stars (FGK dwarfs), or red giant stars: the median ($\pm$ MAD) metallicity obtained for the 48 M dwarfs is [M/H]=+0.09 $\pm$ 0.03, indicating internal consistency in the results.

Stellar luminosities were computed using $K_{\rm s}$-band bolometric corrections \citep{mann2015,mann2016} and Gaia distances \citep{bailerjones2021}, with radii then derived from the luminosity and T$_{\rm eff}$, resulting in values ranging from R=0.22R$_{\odot}$ to 0.53R$_{\odot}$. These radii have a median estimated uncertainty of 6.1\%.
Stellar masses were obtained by combining the luminosities with isochrones having an age of 0.6 Gyr and metallicities of [M/H]=+0.1, with the estimated masses falling in the interval of M$\sim$0.20 -- 0.55 M$_{\odot}$.  
Rotational periods for 28 of the M dwarfs were taken from the literature and used in combination with the masses to calculate Rossby Numbers for this sub-sample of stars. The convective envelope turnover time was computed using the relation in \citet{wright2018_tcz}.

Using the derived effective temperatures and luminosities we investigated radius inflation in the Hyades open cluster M dwarfs.
The M-dwarf radii were compared to predicted radii from different sets of isochrone models that do not consider magnetic fields nor spots:
DARTMOUTH \citep{Dotter2008_dartmouth}, MIST \citep{choi2016_MIST}, and PARSEC \citep{Bressan2012_parsec,nguyen2022}, and were found to be, overall larger than the model isochrones, except for the PARSEC isochrone (which was adjusted to empirical data of stellar radii in eclipsing binaries). 
The median ($\pm$ MAD) radius inflation values for the sample considering MIST and DARTMOUTH models are, respectively, 1.6$\pm$2.3\%, 2.4$\pm$2.3\%.

More detailed comparisons to isochrone models were investigated by dividing the M dwarfs into sub-samples comprised of rapid rotators (P$_{\rm rot} < 5$ days) and slow rotators (P$_{\rm rot} > 5$ days). 
This threshold in rotational periods represents a threshold in Rossby number of $Ro$ = 0.06 for our sample. 
Additional comparisons were done by dividing the sample by mass, in order to segregate partially convective stars (M$ > $0.35M$_{\odot}$) and fully convective stars (M$ < $0.35M$_{\odot}$). Cross-matching the rotational periods with the masses, we find that most of the fully convective M dwarfs with available rotational periods, fall into the rapid rotating group, while the partially convective stars with available rotational periods consist of both rapid (6) and slow (14) rotators.

Taking the MIST isochrone as a baseline the median ($\pm$ MAD) radius inflation obtained for fully- and partially-convective M dwarfs are -0.4$\pm$1.8\% and 2.7$\pm$2.1\% (see Figure \ref{dartmistfig}). This indicates an inhibited or very small radius inflation for the fully-convective M dwarfs, which is in agreement with the predictions from the models that consider magnetic fields \citep{feiden2015}. Also in agreement with the predictions of magnetic models is our finding that there is more significant radius inflation for the more massive and partially convective M dwarfs.

For a DARTMOUTH isochrone, on the contrary, there is more radius inflation for the fully-convective M dwarf regime and less for the partially convective one, as has been identified in other studies in the literature mentioned previously. The median ($\pm$ MAD) radius inflation obtained for fully- and partially-convective M dwarfs are 3.9$\pm$1.5\% and 0.6$\pm$1.6\%. 
In addition, there is no clear difference between partially- and fully-convective M dwarfs, with both samples presenting a similar relation between radius inflation and rotation. 
The results obtained for partially-convective M dwarfs, however, agree with those from MIST isochrone, that, for these stars, radius inflation is dependent on rotation, with partially-convective rapid-rotators (sub-sample composed of just 6 stars) presenting on average higher inflation levels than slow-rotators.  

A comparison with isochrone models that consider different levels of fractional spot coverage \citep{Somers2020_spots} indicates that radius inflation may be explained by stellar spots. Locations in the luminosity-T$_{\rm eff}$ plane covered by the stars in our sample find that 76\% are consistent with fractional spot coverage up to 20\%, while stars exhibiting the highest levels of inflation fall between $\sim$20\% to $\sim$40\% of spot coverage. The median inflation for the sample obtained from SPOTS isochrones is 1.0 $\pm$ 0.6 \%, reaching up to $\sim$5\%, while the maximum inflation obtained from MIST and DARTMOUTH isochrones is, respectively, $\sim$12\% and $\sim$11\%. The inflation pattern of our sample roughly follows the trend expected from SPOTS isochrones, with fully-convective stars presenting constant and moderate inflation levels, and partially-convective M dwarfs presenting a large scatter, with more massive M dwarfs reaching higher levels of inflation. We also found that stars around the threshold that separate fully- from partially-convective M dwarfs present higher inflation levels than their neighbors with lower and higher masses, as expected from SPOTS isochrones.

Finally, ratios of NUV, FUV, and X-ray luminosities to bolometric luminosities compared to fractional photospheric spot coverage, f$_{\rm spot}$, show that the fully-convective sample is both more active and spotted than the partially-convective sample (keeping in mind that the fully-convective stars are composed mostly of rapid rotators), revealing a correlation between activity and spot coverage as derived from SPOTS isochrones.

\acknowledgments

We thank the referee for comments that improved the paper significantly. F.W. acknowledges support from fellowship by Coordena\c c\~ao de Ensino Superior - CAPES. K.C. acknowledges partial support by the National Aeronautics and Space Administration under Grant 18-2ADAP18-0113, issued through the Astrophysics Division of the Science Mission Directorate. K.C. and V.S. acknowledge that their work here is supported,
in part, by the National Science Foundation through NSF grant No. AST-2009507.
D.S. thanks the National Council for Scientific and Technological Development – CNPq.
J.G.F-T gratefully acknowledges the grant support provided by Proyecto Fondecyt Iniciaci\'on No. 11220340, and also from ANID Concurso de Fomento a la Vinculaci\'on Internacional para Instituciones de Investigaci\'on Regionales (Modalidad corta duraci\'on) Proyecto No. FOVI210020, and from the Joint Committee ESO-Government of Chile 2021 (ORP 023/2021), and from Becas Santander Movilidad Internacional Profesores 2022, Banco Santander Chile.
D.M. is supported by the ANID BASAL projects ACE210002 and FB210003 and by Fondecyt Project No. 1220724.

Funding for the Sloan Digital Sky Survey IV has been provided by the Alfred P. Sloan Foundation, the U.S. Department of Energy Office of Science, and the Participating Institutions. SDSS-IV acknowledges support and resources from the Center for High-Performance Computing at the University of Utah. The SDSS web site is www.sdss.org.
SDSS-IV is managed by the Astrophysical Research consortium for the Participating Institutions of the SDSS Collaboration including the Brazilian Participation Group, the Carnegie Institution for Science, Carnegie Mellon University, the Chilean Participation Group, the French Participation Group, Harvard-Smithsonian Center for Astrophysics, Instituto de Astrof\'isica de Canarias, The Johns Hopkins University, 
Kavli Institute for the Physics and Mathematics of the Universe (IPMU) /  University of Tokyo, Lawrence Berkeley National Laboratory, Leibniz Institut f\"ur Astrophysik Potsdam (AIP),  Max-Planck-Institut f\"ur Astronomie (MPIA Heidelberg), Max-Planck-Institut f\"ur Astrophysik (MPA Garching), Max-Planck-Institut f\"ur Extraterrestrische Physik (MPE), National Astronomical Observatory of China, New Mexico State University, New York University, University of Notre Dame, Observat\'orio Nacional / MCTI, The Ohio State University, Pennsylvania State University, Shanghai Astronomical Observatory, United Kingdom Participation Group,
Universidad Nacional Aut\'onoma de M\'exico, University of Arizona, University of Colorado Boulder, University of Oxford, University of Portsmouth, University of Utah, University of Virginia, University of Washington, University of Wisconsin, Vanderbilt University, and Yale University.

\facility {Sloan}

\software{Turbospectrum (\citealt{plez2012_turbospectrum}), Bacchus (\citealt{masseron2016_bacchus}), Matplotlib (\citealt{Hunter2007_matplotlib}), Numpy (\citealt{harris2020_numpy}).}

\startlongtable
\begin{longrotatetable}
\begin{deluxetable*}{lcccccccccccccccc}
\tablenum{1}
\label{stellartable}
\tabletypesize{\scriptsize}
\tablecaption{Astrometric, photometric and spectroscopic parameters}
\tablewidth{0pt}
\tablehead{
\colhead{APOGEE ID} &
\colhead{G} &
\colhead{$G_{\rm BP}$} &
\colhead{$G_{\rm RP}$} &
\colhead{H} &
\colhead{$K_{\rm s}$} &
\colhead{RV} &
\colhead{d} &
\colhead{$T_{\rm eff}$} &
\colhead{$\log{g}$} &
\colhead{[M/H]} &
\colhead{A(O)} &
\colhead{V$\sin{i}$} &
\colhead{BC$_{\rm K_{\rm s}}$} &
\colhead{L$_{\rm bol}$} &
\colhead{S/N} & \\
\colhead{...} &
\colhead{...} &
\colhead{...} &
\colhead{...} &
\colhead{...} &
\colhead{...} &
\colhead{km/s} &
\colhead{pc} &
\colhead{K} &
\colhead{...} &
\colhead{...} &
\colhead{...} &
\colhead{km/s} &
\colhead{...} &
\colhead{$10^{31}$} &
\colhead{...}
}
\startdata
2M03534647+1323312 & 14.81 & 16.70 & 13.50 & 10.98 & 10.68 & 36.23 & 41.22$\pm$0.06 & 3124 & 5.06 & 0.13 & 8.84$\pm$0.02 & 9.4$\pm$1.40 & 2.83 & 2.0$\pm$0.1 & 135 \\
2M05011599+1608409 & 15.80 & 17.69 & 14.49 & 12.00 & 11.71 & 43.43 & 65.25$\pm$0.16 & 3137 & 5.00 & 0.09 & 8.79$\pm$0.02 & 5.5$\pm$1.60 & 2.89 & 1.9$\pm$0.1 & 77 \\
2M04242092+1910509 & 15.13 & 16.92 & 13.86 & 11.42 & 11.13 & 39.73 & 46.99$\pm$0.06 & 3167 & 5.01 & 0.10 & 8.79$\pm$0.03 & 18.6$\pm$0.70 & 2.86 & 1.7$\pm$0.1 & 112 \\
2M04335544+1822507 & 14.97 & 16.76 & 13.70 & 11.24 & 10.97 & 40.60 & 46.20$\pm$0.06 & 3215 & 5.07 & 0.10 & 8.80$\pm$0.02 & 10.5$\pm$0.80 & 2.79 & 2.0$\pm$0.1 & 194 \\
2M03581434+1237408 & 15.08 & 16.98 & 13.78 & 11.26 & 10.99 & 37.36 & 45.45$\pm$0.09 & 3219 & 5.07 & 0.08 & 8.80$\pm$0.03 & 7.8$\pm$1.40 & 2.81 & 1.9$\pm$0.1 & 113 \\
2M04183382+1821529 & 14.59 & 16.35 & 13.32 & 10.92 & 10.60 & 39.87 & 45.71$\pm$0.08 & 3240 & 4.93 & 0.06 & 8.76$\pm$0.01 & 7.8$\pm$1.30 & 2.76 & 2.8$\pm$0.1 & 118 \\
2M04493656+1701593 & 14.76 & 16.54 & 13.49 & 11.01 & 10.75 & 42.03 & 50.26$\pm$0.06 & 3247 & 4.99 & 0.06 & 8.77$\pm$0.02 & 9.8$\pm$1.50 & 2.81 & 2.8$\pm$0.1 & 98 \\
2M04001556+1924367 & 14.68 & 16.42 & 13.42 & 10.97 & 10.70 & 36.93 & 47.96$\pm$0.07 & 3262 & 5.00 & 0.07 & 8.76$\pm$0.02 & 8.4$\pm$1.30 & 2.77 & 2.8$\pm$0.1 & 73 \\
2M04175160+1513378 & 14.22 & 15.86 & 12.98 & 10.64 & 10.36 & 38.91 & 47.19$\pm$0.06 & 3262 & 4.86 & 0.09 & 8.77$\pm$0.03 & 5.5$\pm$0.90 & 2.77 & 3.8$\pm$0.1 & 140 \\
2M04270314+2406159 & 14.59 & 16.38 & 13.32 & 10.88 & 10.57 & 38.91 & 47.06$\pm$0.06 & 3263 & 5.02 & 0.13 & 8.82$\pm$0.02 & 5.1$\pm$2.50 & 2.83 & 2.9$\pm$0.1 & 118 \\
2M04301702+2622264 & 14.94 & 16.74 & 13.67 & 11.16 & 10.91 & 38.85 & 54.75$\pm$0.10 & 3268 & 5.02 & 0.07 & 8.78$\pm$0.03 & 10.3$\pm$1.10 & 2.84 & 2.9$\pm$0.1 & 108 \\
2M03390711+2025267 & 13.60 & 15.28 & 12.36 & 9.97 & 9.71 & 33.71 & 35.68$\pm$0.03 & 3273 & 5.00 & 0.15 & 8.86$\pm$0.01 & 6.5$\pm$0.70 & 2.77 & 3.9$\pm$0.1 & 274 \\
2M04334280+1845592 & 13.96 & 15.65 & 12.70 & 10.32 & 10.05 & 40.30 & 41.28$\pm$0.03 & 3290 & 5.02 & 0.08 & 8.78$\pm$0.01 & 4.4$\pm$0.90 & 2.78 & 3.8$\pm$0.1 & 324 \\
2M04551484+2121505 & 15.24 & 17.00 & 13.98 & 11.55 & 11.26 & 42.13 & 60.00$\pm$0.10 & 3300 & 5.04 & 0.13 & 8.83$\pm$0.02 & 13.0$\pm$1.70 & 2.82 & 2.5$\pm$0.1 & 98 \\
2M04130560+1514520 & 14.36 & 16.06 & 13.10 & 10.73 & 10.44 & 38.76 & 44.62$\pm$0.06 & 3314 & 4.98 & 0.04 & 8.77$\pm$0.01 & 10.1$\pm$1.40 & 2.78 & 3.1$\pm$0.1 & 118 \\
2M04190307+1932395 & 13.95 & 15.65 & 12.70 & 10.29 & 10.01 & 39.06 & 43.07$\pm$0.04 & 3318 & 4.94 & 0.12 & 8.82$\pm$0.02 & 5.8$\pm$0.70 & 2.77 & 4.3$\pm$0.1 & 119 \\
2M04575778+1427071 & 14.37 & 16.06 & 13.12 & 10.71 & 10.46 & 42.64 & 48.39$\pm$0.07 & 3343 & 5.02 & 0.06 & 8.76$\pm$0.02 & 8.9$\pm$1.10 & 2.78 & 3.6$\pm$0.1 & 107 \\
2M04235738+2210537 & 13.82 & 15.44 & 12.59 & 10.25 & 10.00 & 39.23 & 42.96$\pm$0.04 & 3346 & 4.94 & 0.07 & 8.78$\pm$0.01 & 4.1$\pm$1.00 & 2.76 & 4.4$\pm$0.1 & 128 \\
2M04402825+1805160 & 13.66 & 15.33 & 12.41 & 10.03 & 9.76 & 40.65 & 38.89$\pm$0.03 & 3360 & 4.99 & 0.06 & 8.78$\pm$0.02 & 2.2$\pm$2.00 & 2.77 & 4.4$\pm$0.1 & 188 \\
2M04561668+2122019 & 14.02 & 15.59 & 12.81 & 10.50 & 10.25 & 41.74 & 52.96$\pm$0.06 & 3392 & 4.87 & 0.05 & 8.76$\pm$0.01 & 4.4$\pm$1.60 & 2.70 & 5.6$\pm$0.2 & 152 \\
2M04202761+1853499 & 13.99 & 15.56 & 12.77 & 10.48 & 10.22 & 39.26 & 50.69$\pm$0.07 & 3407 & 4.94 & 0.09 & 8.80$\pm$0.02 & 6.0$\pm$0.40 & 2.75 & 5.0$\pm$0.2 & 148 \\
2M04425217+2320224 & 13.74 & 15.25 & 12.54 & 10.21 & 10.01 & 40.57 & 53.85$\pm$0.05 & 3407 & 4.83 & 0.14 & 8.84$\pm$0.01 & 6.6$\pm$0.70 & 2.73 & 7.0$\pm$0.2 & 152 \\
2M04425849+2036174 & 13.77 & 15.26 & 12.57 & 10.35 & 10.08 & 41.34 & 51.34$\pm$0.05 & 3407 & 4.82 & 0.04 & 8.74$\pm$0.02 & 4.4$\pm$0.70 & 2.72 & 6.0$\pm$0.2 & 197 \\
2M04311101+1623452 & 13.79 & 15.34 & 12.58 & 10.28 & 10.04 & 40.37 & 49.09$\pm$0.05 & 3420 & 4.93 & 0.04 & 8.75$\pm$0.02 & 5.3$\pm$1.20 & 2.80 & 5.3$\pm$0.2 & 160 \\
2M04574644+1628202 & 14.42 & 16.03 & 13.19 & 10.85 & 10.59 & 43.40 & 65.52$\pm$0.16 & 3426 & 4.91 & 0.05 & 8.78$\pm$0.02 & 7.4$\pm$0.60 & 2.75 & 5.9$\pm$0.2 & 126 \\
2M04335669+1652087 & 13.74 & 15.29 & 12.53 & 10.23 & 9.96 & 40.49 & 46.42$\pm$0.04 & 3433 & 4.99 & 0.10 & 8.80$\pm$0.02 & 5.6$\pm$1.70 & 2.74 & 5.4$\pm$0.2 & 239 \\
2M03434706+2051363 & 13.31 & 14.81 & 12.12 & 9.83 & 9.61 & 35.04 & 44.85$\pm$0.05 & 3439 & 4.88 & 0.11 & 8.80$\pm$0.02 & 4.4$\pm$1.00 & 2.72 & 7.1$\pm$0.2 & 255 \\
2M04322897+1754166 & 14.14 & 15.71 & 12.93 & 10.64 & 10.38 & 27.32 & 56.57$\pm$0.07 & 3442 & 5.01 & 0.03 & 8.75$\pm$0.02 & 9.6$\pm$0.70 & 2.74 & 5.5$\pm$0.2 & 174 \\
2M04363080+1905273 & 14.02 & 14.96 & 12.24 & 9.97 & 9.71 & 40.65 & 59.13$\pm$0.23 & 3449 & 4.93 & 0.10 & 8.80$\pm$0.01 & 5.0$\pm$1.30 & 2.73 & 11.2$\pm$0.4 & 313 \\
2M04501665+2037330 & 13.13 & 14.62 & 11.94 & 9.63 & 9.39 & 42.06 & 45.23$\pm$0.04 & 3456 & 4.79 & 0.11 & 8.81$\pm$0.03 & 18.6$\pm$0.90 & 2.72 & 8.8$\pm$0.3 & 263 \\
2M04421880+1741383 & 13.76 & 15.28 & 12.56 & 10.28 & 10.02 & 41.50 & 51.66$\pm$0.05 & 3462 & 4.87 & 0.04 & 8.75$\pm$0.02 & 5.6$\pm$0.40 & 2.73 & 6.4$\pm$0.2 & 170 \\
2M04370517+2043054 & 13.99 & 15.50 & 12.80 & 10.52 & 10.28 & 40.64 & 57.70$\pm$0.07 & 3483 & 4.92 & 0.03 & 8.75$\pm$0.01 & 6.2$\pm$0.90 & 2.83 & 5.7$\pm$0.2 & 110 \\
2M03413689+2320374 & 12.74 & 14.11 & 11.59 & 9.42 & 9.16 & 34.28 & 41.22$\pm$0.06 & 3505 & 4.82 & 0.13 & 8.82$\pm$0.03 & 6.1$\pm$0.90 & 2.69 & 9.3$\pm$0.3 & 271 \\
2M04172811+1454038 & 13.26 & 14.68 & 12.08 & 9.87 & 9.62 & 39.32 & 49.16$\pm$0.04 & 3522 & 4.87 & 0.03 & 8.75$\pm$0.01 & 10.3$\pm$0.70 & 2.71 & 8.5$\pm$0.3 & 253 \\
2M04192976+2145141 & 12.85 & 14.23 & 11.69 & 9.44 & 9.22 & 39.25 & 47.52$\pm$0.05 & 3523 & 4.72 & 0.11 & 8.82$\pm$0.03 & 19.1$\pm$0.30 & 2.70 & 11.6$\pm$0.4 & 236 \\
2M04293738+2140072 & 13.01 & 14.46 & 11.83 & 9.64 & 9.41 & 39.93 & 42.60$\pm$0.04 & 3534 & 4.87 & 0.07 & 8.79$\pm$0.01 & 6.0$\pm$0.80 & 2.70 & 7.8$\pm$0.3 & 117 \\
2M03585452+2513113 & 13.00 & 14.30 & 11.87 & 9.76 & 9.50 & 35.58 & 49.38$\pm$0.10 & 3547 & 4.78 & 0.08 & 8.77$\pm$0.01 & 5.0$\pm$0.60 & 2.66 & 10.0$\pm$0.3 & 311 \\
2M04350330+2451146 & 13.48 & 14.81 & 12.34 & 10.18 & 9.92 & 33.28 & 59.33$\pm$0.08 & 3552 & 4.83 & 0.15 & 8.82$\pm$0.02 & 6.0$\pm$0.50 & 2.69 & 9.6$\pm$0.3 & 182 \\
2M04544410+1940513 & 12.83 & 14.17 & 11.68 & 9.49 & 9.28 & 42.28 & 45.35$\pm$0.04 & 3579 & 4.82 & 0.10 & 8.80$\pm$0.00 & 5.4$\pm$0.80 & 2.68 & 10.2$\pm$0.3 & 212 \\
2M04290099+1840254 & 12.25 & 13.53 & 11.13 & 8.95 & 8.69 & 41.20 & 43.20$\pm$0.04 & 3585 & 4.69 & 0.19 & 8.90$\pm$0.03 & 18.4$\pm$0.40 & 2.67 & 16.0$\pm$0.5 & 549 \\
2M04401271+1917099 & 12.55 & 13.81 & 11.44 & 9.36 & 9.12 & 40.84 & 44.88$\pm$0.05 & 3590 & 4.74 & 0.09 & 8.78$\pm$0.02 & 5.1$\pm$1.30 & 2.66 & 11.8$\pm$0.4 & 181 \\
2M04385471+1910560 & 12.90 & 14.25 & 11.75 & 9.51 & 9.27 & 40.33 & 49.80$\pm$0.05 & 3636 & 4.84 & 0.10 & 8.83$\pm$0.01 & 25.0$\pm$1.75 & 2.69 & 12.3$\pm$0.4 & 294 \\
2M04491107+1742557 & 13.00 & 14.22 & 11.90 & 9.84 & 9.60 & 42.34 & 58.38$\pm$0.06 & 3642 & 4.76 & 0.07 & 8.74$\pm$0.00 & 4.3$\pm$1.55 & 2.65 & 12.9$\pm$0.4 & 124 \\
2M03415547+1845359 & 12.23 & 13.45 & 11.14 & 9.11 & 8.87 & 34.77 & 41.21$\pm$0.04 & 3642 & 4.78 & 0.10 & 8.79$\pm$0.02 & 4.3$\pm$0.40 & 2.65 & 12.6$\pm$0.4 & 330 \\
2M04363893+1836567 & 12.30 & 13.52 & 11.20 & 9.15 & 8.94 & 40.99 & 42.00$\pm$0.03 & 3646 & 4.76 & 0.06 & 8.75$\pm$0.01 & 3.7$\pm$0.50 & 2.65 & 12.3$\pm$0.4 & 519 \\
2M04295572+1654506 & 11.97 & 13.10 & 10.90 & 8.90 & 8.65 & 40.38 & 43.98$\pm$0.04 & 3709 & 4.67 & 0.14 & 8.84$\pm$0.03 & 5.7$\pm$0.30 & 2.60 & 18.3$\pm$0.6 & 514 \\
2M04291097+2614484 & 12.06 & 13.15 & 11.01 & 9.04 & 8.83 & 38.26 & 47.80$\pm$0.04 & 3724 & 4.67 & 0.16 & 8.84$\pm$0.02 & 6.0$\pm$0.30 & 2.60 & 18.5$\pm$0.6 & 297 \\
2M03591417+2202380 & 12.07 & 13.26 & 10.98 & 8.91 & 8.70 & 35.20 & 40.38$\pm$0.04 & 3744 & 4.82 & 0.05 & 8.79$\pm$0.00 & 5.2$\pm$1.70 & 2.64 & 14.2$\pm$0.5 & 226 \\
\enddata
\tablenotetext{}{\tablenotetext{}{The estimated uncertainties in the derived parameters are presented in Section 3.4.}}
\end{deluxetable*}
\end{longrotatetable}

\startlongtable
\begin{deluxetable*}{lcccc}
\tablenum{2}
\label{Spectral_Lines}
\tabletypesize{\scriptsize}
\tablecaption{Spectral Lines}
\tablewidth{0pt}
\tablehead{
\colhead{Species} &
\colhead{$\lambda$} &
\colhead{$\chi_{\rm exc}$} &
\colhead{$\log{gf}$} & \\
\colhead{...} &
\colhead{(\r{A})} &
\colhead{(eV)} &
\colhead{...} &
}
\startdata
OH & 15266.168 & 0.210 & -5.500 \\
OH & 15281.055 & 0.205 & -5.453 \\
OH & 15372.539 & 1.053 & -5.102 \\
OH & 15391.205 & 0.494 & -5.512 \\
OH & 15407.294 & 0.255 & -5.435 \\
OH & 15409.170 & 0.255 & -5.435 \\
OH & 15505.324 & 0.515 & -5.378 \\
OH & 15505.746 & 0.515 & -5.378 \\
OH & 15558.017 & 0.304 & -5.375 \\
OH & 15560.245 & 0.304 & -5.375 \\
OH & 15565.815 & 0.898 & -5.386 \\
OH & 15566.000 & 0.898 & -5.386 \\
OH & 15568.782 & 0.299 & -5.337 \\
OH & 15572.084 & 0.300 & -5.337 \\
OH & 15627.289 & 0.886 & -5.514 \\
OH & 16052.766 & 0.639 & -4.976 \\
OH & 16061.702 & 0.476 & -5.222 \\
OH & 16534.582 & 0.781 & -4.806 \\
OH & 16871.893 & 0.763 & -5.056 \\
OH & 16872.277 & 0.759 & -5.032 \\
H$_{2}$O & 15255-15261 & ... & ... \\
H$_{2}$O & 15269-15272 & ... & ... \\
H$_{2}$O & 15314-15319 & ... & ... \\
H$_{2}$O & 15352-15355 & ... & ... \\
H$_{2}$O & 15359-15362 & ... & ... \\
H$_{2}$O & 15446-15450 & ... & ... \\
H$_{2}$O & 15455-15457 & ... & ... \\
H$_{2}$O & 15459-15463 & ... & ... \\
H$_{2}$O & 15502-15505 & ... & ... \\
\enddata
\end{deluxetable*}

\startlongtable
\begin{longrotatetable}
\begin{deluxetable*}{lccccccccccccccccccc}
\tablenum{3}
\label{rotationaltable}
\tabletypesize{\scriptsize}
\tablecaption{Masses, Radii, Rotational Periods and Activity Data}
\tablewidth{0pt}
\tablehead{
\colhead{APOGEE ID} &
\colhead{R$_{*}$} &
\colhead{P$_{\rm rot}$} &
\colhead{M} &
\colhead{M} &
\colhead{M} &
\colhead{$<$M$>$} &
\colhead{R$_{\rm frac}$} &
\colhead{R$_{\rm frac}$} &
\colhead{R$_{\rm frac}$} &
\colhead{$\tau_{\rm cz}$} &
\colhead{$\tau_{\rm cz}$} &
\colhead{$\tau_{\rm cz}$} &
\colhead{Ro} &
\colhead{Ro} &
\colhead{Ro} &
\colhead{L$_{\rm nuv}$} &
\colhead{L$_{\rm fuv}$} &
\colhead{L$_{\rm x}$} & \\
\colhead{...} &
\colhead{R$_{\odot}$} &
\colhead{day} &
\colhead{M$_{\odot}$} &
\colhead{M$_{\odot}$} &
\colhead{M$_{\odot}$} &
\colhead{M$_{\odot}$} &
\colhead{...} &
\colhead{...} &
\colhead{...} &
\colhead{...} &
\colhead{...} &
\colhead{...} &
\colhead{...} &
\colhead{...} &
\colhead{...} &
\colhead{$10^{27}$} &
\colhead{$10^{27}$} &
\colhead{$10^{27}$} \\
\colhead{...} &
\colhead{...} &
\colhead{...} &
\colhead{MIS} &
\colhead{DAR} &
\colhead{SPO} &
\colhead{...} &
\colhead{MIS} &
\colhead{DAR} &
\colhead{SPO} &
\colhead{MIS} &
\colhead{DAR} &
\colhead{SPO} &
\colhead{MIS} &
\colhead{DAR} &
\colhead{SPO} &
\colhead{...} &
\colhead{...} &
\colhead{...}}
\startdata
2M04192976+2145141 & 0.47$\pm$0.03 & 0.6 & 0.47 & 0.48 & 0.50 & 0.483$\pm$0.011 & 0.08 & 0.05 & 0.03 & 49.2 & 48.2 & 45.7 & 0.01 & 0.01 & 0.01 & ... & ... & 94.5  \\
2M04242092+1910509 & 0.22$\pm$0.01 & 0.6 & 0.21 & 0.19 & 0.20 & 0.198$\pm$0.006 & -0.03 & 0.06 & 0.01 & 108.3 & 113.7 & 111.0 & 0.01 & 0.01 & 0.01 & ... & ... & 5.7  \\
2M03581434+1237408 & 0.23$\pm$0.01 & 0.9 & 0.22 & 0.20 & 0.21 & 0.209$\pm$0.006 & -0.04 & 0.04 & 0.00 & 104.7 & 109.6 & 107.7 & 0.01 & 0.01 & 0.01 & ... & ... & 10.8  \\
2M04385471+1910560 & 0.45$\pm$0.03 & 0.9 & 0.48 & 0.49 & 0.49 & 0.487$\pm$0.005 & 0.02 & 0.00 & 0.00 & 47.9 & 47.1 & 46.3 & 0.02 & 0.02 & 0.02 & 30.4 & 4.3 & ...  \\
2M04501665+2037330 & 0.42$\pm$0.03 & 1.0 & 0.43 & 0.44 & 0.45 & 0.44$\pm$0.01 & 0.07 & 0.04 & 0.02 & 55.4 & 54.2 & 51.7 & 0.02 & 0.02 & 0.02 & 28.2 & 7.3 & ...  \\
2M05011599+1608409 & 0.24$\pm$0.02 & 1.4 & 0.22 & 0.20 & 0.21 & 0.21$\pm$0.007 & 0.01 & 0.09 & 0.02 & 105.1 & 110.1 & 105.6 & 0.01 & 0.01 & 0.01 & 8.7 & 14.9 & ...  \\
2M04551484+2121505 & 0.25$\pm$0.02 & 1.4 & 0.25 & 0.24 & 0.24 & 0.242$\pm$0.007 & -0.06 & 0.01 & 0.00 & 93.8 & 97.8 & 98.2 & 0.02 & 0.01 & 0.01 & 7.5 & 9.0 & 36.4  \\
2M04493656+1701593 & 0.27$\pm$0.02 & 1.4 & 0.27 & 0.26 & 0.27 & 0.263$\pm$0.005 & 0.00 & 0.05 & 0.01 & 89.4 & 92.5 & 89.8 & 0.02 & 0.02 & 0.02 & 9.4 & 6.9 & 26.1  \\
2M04335544+1822507 & 0.23$\pm$0.02 & 1.5 & 0.22 & 0.21 & 0.22 & 0.217$\pm$0.006 & -0.03 & 0.05 & 0.00 & 102.0 & 107.0 & 104.5 & 0.01 & 0.01 & 0.01 & 8.1 & ... & ...  \\
2M04290099+1840254 & 0.53$\pm$0.03 & 1.5 & 0.52 & 0.53 & 0.55 & 0.533$\pm$0.011 & 0.10 & 0.08 & 0.05 & 42.7 & 42.3 & 39.8 & 0.04 & 0.04 & 0.04 & ... & ... & ...  \\
2M04322897+1754166 & 0.34$\pm$0.02 & 1.7 & 0.36 & 0.36 & 0.36 & 0.358$\pm$0.001 & -0.01 & -0.01 & -0.01 & 68.1 & 68.2 & 67.6 & 0.02 & 0.02 & 0.03 & 15.0 & ... & ...  \\
2M04001556+1924367 & 0.27$\pm$0.02 & 1.9 & 0.27 & 0.26 & 0.26 & 0.261$\pm$0.004 & -0.01 & 0.04 & 0.01 & 89.5 & 92.6 & 91.2 & 0.02 & 0.02 & 0.02 & ... & ... & 32.6  \\
2M04130560+1514520 & 0.27$\pm$0.02 & 1.9 & 0.28 & 0.27 & 0.27 & 0.27$\pm$0.006 & -0.04 & 0.01 & 0.01 & 86.6 & 89.6 & 89.9 & 0.02 & 0.02 & 0.02 & ... & 3.5 & ...  \\
2M04575778+1427071 & 0.29$\pm$0.02 & 2.0 & 0.30 & 0.29 & 0.29 & 0.292$\pm$0.003 & -0.04 & 0.01 & 0.00 & 81.9 & 83.6 & 83.5 & 0.03 & 0.02 & 0.02 & ... & ... & ...  \\
2M04183382+1821529 & 0.27$\pm$0.02 & 2.3 & 0.27 & 0.26 & 0.26 & 0.262$\pm$0.005 & 0.00 & 0.05 & 0.02 & 89.6 & 92.7 & 90.2 & 0.03 & 0.03 & 0.03 & 3.2 & ... & 6  \\
2M04172811+1454038 & 0.40$\pm$0.02 & 2.4 & 0.42 & 0.43 & 0.44 & 0.43$\pm$0.005 & 0.03 & 0.00 & 0.00 & 56.4 & 55.1 & 54.4 & 0.04 & 0.04 & 0.04 & ... & ... & 45.2  \\
2M04175160+1513378 & 0.31$\pm$0.02 & 3.6 & 0.30 & 0.30 & 0.32 & 0.307$\pm$0.009 & 0.02 & 0.07 & 0.01 & 80.0 & 81.3 & 76.3 & 0.05 & 0.04 & 0.05 & ... & ... & ...  \\
2M04334280+1845592 & 0.31$\pm$0.02 & 4.8 & 0.31 & 0.30 & 0.31 & 0.307$\pm$0.006 & 0.00 & 0.05 & 0.01 & 79.6 & 80.8 & 77.6 & 0.06 & 0.06 & 0.06 & 28.5 & ... & ...  \\
2M04425849+2036174 & 0.36$\pm$0.02 & 10.5 & 0.37 & 0.37 & 0.38 & 0.374$\pm$0.005 & 0.03 & 0.02 & 0.01 & 65.7 & 65.3 & 63.7 & 0.16 & 0.16 & 0.17 & 3.9 & 5.7 & ...  \\
2M04370517+2043054 & 0.33$\pm$0.02 & 11.9 & 0.36 & 0.36 & ... & 0.364$\pm$0.001 & -0.03 & -0.03 & ... & 67.0 & 66.9 & ... & 0.18 & 0.18 & ... & 10.0 & ... & ...  \\
2M04421880+1741383 & 0.36$\pm$0.02 & 12.2 & 0.38 & 0.38 & 0.38 & 0.382$\pm$0.002 & 0.01 & 0.00 & 0.00 & 63.9 & 63.3 & 62.9 & 0.19 & 0.19 & 0.19 & 9.4 & 4.6 & ...  \\
2M03434706+2051363 & 0.38$\pm$0.02 & 12.3 & 0.39 & 0.40 & 0.41 & 0.402$\pm$0.006 & 0.04 & 0.02 & 0.01 & 61.2 & 60.0 & 58.6 & 0.20 & 0.21 & 0.21 & ... & ... & 25.1  \\
2M04363080+1905273 & 0.48$\pm$0.03 & 13.1 & 0.47 & 0.47 & 0.50 & 0.479$\pm$0.015 & 0.12 & 0.09 & 0.05 & 50.0 & 49.0 & 45.5 & 0.26 & 0.27 & 0.29 & 10.0 & 11.6 & ...  \\
2M04295572+1654506 & 0.53$\pm$0.03 & 14.5 & 0.54 & 0.55 & 0.56 & 0.551$\pm$0.008 & 0.06 & 0.04 & 0.03 & 40.3 & 40.2 & 38.5 & 0.36 & 0.36 & 0.38 & 22.5 & 3.3 & 58.7  \\
2M04291097+2614484 & 0.53$\pm$0.03 & 18.0 & 0.55 & 0.55 & 0.56 & 0.552$\pm$0.007 & 0.05 & 0.03 & 0.02 & 40.1 & 40.0 & 38.4 & 0.45 & 0.45 & 0.47 & ... & ... & 22.9  \\
2M04202761+1853499 & 0.33$\pm$0.02 & 20.3 & 0.35 & 0.35 & 0.35 & 0.346$\pm$0.001 & 0.00 & 0.00 & 0.01 & 70.5 & 70.5 & 70.4 & 0.29 & 0.29 & 0.29 & ... & ... & 13  \\
2M03591417+2202380 & 0.46$\pm$0.03 & 20.7 & 0.50 & 0.51 & ... & 0.507$\pm$0.002 & -0.01 & -0.03 & ... & 44.9 & 44.3 & ... & 0.46 & 0.47 & ... & 6.7 & 4.0 & 2.1  \\
2M04401271+1917099 & 0.45$\pm$0.03 & 21.5 & 0.47 & 0.48 & 0.49 & 0.482$\pm$0.007 & 0.04 & 0.01 & 0.01 & 48.8 & 47.9 & 46.5 & 0.44 & 0.45 & 0.46 & 6.6 & 4.6 & ...  \\
2M04425217+2320224 & 0.39$\pm$0.02 & 21.8 & 0.39 & 0.40 & 0.42 & 0.402$\pm$0.01 & 0.06 & 0.04 & 0.01 & 61.5 & 60.3 & 57.6 & 0.35 & 0.36 & 0.38 & ... & ... & 29.9  \\
2M04363893+1836567 & 0.45$\pm$0.03 & 21.9 & 0.48 & 0.49 & 0.49 & 0.488$\pm$0.005 & 0.02 & -0.01 & 0.00 & 47.8 & 47.0 & 46.3 & 0.46 & 0.47 & 0.47 & 7.5 & 5.6 & 5.5  \\
2M04235738+2210537 & 0.32$\pm$0.02 & 21.9 & 0.33 & 0.32 & 0.33 & 0.328$\pm$0.003 & 0.01 & 0.02 & 0.01 & 74.4 & 75.1 & 73.4 & 0.29 & 0.29 & 0.30 & 6.8 & ... & 15.5  \\
2M04270314+2406159 & 0.27$\pm$0.02 & 28.0 & 0.27 & 0.26 & 0.27 & 0.265$\pm$0.005 & -0.01 & 0.04 & 0.01 & 88.5 & 91.7 & 89.8 & 0.32 & 0.31 & 0.31 & 11.1 & ... & 37.7  \\
2M04335669+1652087 & 0.33$\pm$0.02 & 31.0 & 0.36 & 0.36 & 0.36 & 0.356$\pm$0.001 & -0.01 & 0.00 & 0.00 & 68.6 & 68.6 & 68.4 & 0.45 & 0.45 & 0.45 & 2.8 & 0.4 & ...  \\
2M04350330+2451146 & 0.42$\pm$0.02 & 59.0 & 0.44 & 0.45 & 0.46 & 0.449$\pm$0.006 & 0.03 & 0.00 & 0.00 & 53.5 & 52.3 & 51.3 & 1.10 & 1.13 & 1.15 & ... & ... & ...  \\
2M04311101+1623452 & 0.34$\pm$0.02 & 68.4 & 0.35 & 0.35 & 0.35 & 0.354$\pm$0.001 & 0.00 & 0.00 & 0.01 & 68.9 & 68.9 & 68.8 & 0.99 & 0.99 & 0.99 & ... & ... & 9  \\
2M03390711+2025267 & 0.31$\pm$0.02 & ... & 0.31 & 0.31 & 0.32 & 0.313$\pm$0.009 & 0.02 & 0.06 & 0.01 & 79.1 & 79.6 & 75.1 & ... & ... & ... & 6.7 & 2.4 & 14.4  \\
2M03413689+2320374 & 0.42$\pm$0.03 & ... & 0.44 & 0.44 & 0.45 & 0.445$\pm$0.007 & 0.05 & 0.02 & 0.02 & 54.2 & 53.1 & 51.5 & ... & ... & ... & ... & ... & 19.5  \\
2M03415547+1845359 & 0.46$\pm$0.03 & ... & 0.49 & 0.49 & 0.50 & 0.492$\pm$0.005 & 0.03 & 0.00 & 0.00 & 47.3 & 46.5 & 45.6 & ... & ... & ... & 13.2 & 4.8 & 20.4  \\
2M03534647+1323312 & 0.25$\pm$0.02 & ... & 0.23 & 0.21 & 0.23 & 0.222$\pm$0.009 & 0.02 & 0.11 & 0.02 & 101.9 & 106.9 & 100.1 & ... & ... & ... & ... & ... & 21.5  \\
2M03585452+2513113 & 0.43$\pm$0.03 & ... & 0.45 & 0.46 & 0.47 & 0.457$\pm$0.008 & 0.04 & 0.01 & 0.01 & 52.4 & 51.3 & 49.8 & ... & ... & ... & 5.4 & 6.1 & 42.9  \\
2M04190307+1932395 & 0.32$\pm$0.02 & ... & 0.33 & 0.32 & 0.33 & 0.327$\pm$0.004 & 0.02 & 0.04 & 0.02 & 74.8 & 75.6 & 73.6 & ... & ... & ... & ... & ... & 12.5  \\
2M04293738+2140072 & 0.38$\pm$0.02 & ... & 0.41 & 0.42 & ... & 0.413$\pm$0.004 & 0.00 & -0.02 & ... & 58.6 & 57.3 & ... & ... & ... & ... & 5.0 & ... & 19.5  \\
2M04301702+2622264 & 0.27$\pm$0.02 & ... & 0.27 & 0.26 & 0.26 & 0.262$\pm$0.004 & -0.02 & 0.03 & 0.01 & 89.3 & 92.4 & 91.0 & ... & ... & ... & 8.2 & ... & 10.1  \\
2M04402825+1805160 & 0.32$\pm$0.02 & ... & 0.33 & 0.33 & 0.33 & 0.329$\pm$0.001 & 0.00 & 0.02 & 0.01 & 74.1 & 74.6 & 73.9 & ... & ... & ... & 9.8 & 4.0 & ...  \\
2M04491107+1742557 & 0.46$\pm$0.03 & ... & 0.49 & 0.49 & 0.50 & 0.495$\pm$0.006 & 0.03 & 0.00 & 0.01 & 46.9 & 46.1 & 45.1 & ... & ... & ... & ... & ... & ...  \\
2M04544410+1940513 & 0.42$\pm$0.02 & ... & 0.45 & 0.46 & 0.46 & 0.458$\pm$0.005 & 0.03 & 0.00 & 0.00 & 52.0 & 50.9 & 50.3 & ... & ... & ... & ... & ... & ...  \\
2M04561668+2122019 & 0.35$\pm$0.02 & ... & 0.36 & 0.36 & 0.37 & 0.363$\pm$0.003 & 0.02 & 0.02 & 0.01 & 67.4 & 67.4 & 66.2 & ... & ... & ... & ... & ... & ...  \\
2M04574644+1628202 & 0.35$\pm$0.02 & ... & 0.37 & 0.37 & 0.37 & 0.371$\pm$0.002 & 0.01 & 0.01 & 0.01 & 65.9 & 65.6 & 64.8 & ... & ... & ... & 312.4 & 209.6 & ...  \\
\enddata
\tablenotetext{}{\tablenotetext{}{Measurements associated with MIST, DARTMOUTH and SPOTS are labeled respectively as ``MIS", ``DAR" and ``SPO". The estimated uncertainties in the derived  masses and radii are presented in Section 3.4.}}
\end{deluxetable*}
\end{longrotatetable}

\bibliographystyle{yahapj}
\bibliography{references.bib}

\end{document}